\documentclass[pra,aps,amssymb,twocolumn,10pt,floatfix,showpacs]{revtex4-1}
\usepackage{bbm}
\usepackage{amsmath}
\usepackage{amssymb}
\usepackage{mathtools}
\usepackage{enumerate}
\usepackage{color,graphicx}
\usepackage{enumitem}
\usepackage{makecell}

\newcommand{\ket}[1]{|#1\rangle}
\newcommand*{\bra}[1]{\langle#1|}
\newcommand{\sx}{s_{\mathbf{x}}}
\newcommand{\sutwo}{\text{SU(2)}}

\newcommand{\uone}{\text{U(1)}}
\newcommand{\ztwo}{\mathbb{Z}_2}
\newcommand{\psione}{\ket{\psi_1}}
\newcommand{\psitwo}{\ket{\psi_{2}}}
\newcommand{\sbar}{\bar{s}}
\newcommand{\amap}{\mathcal{A}}
\newcommand{\ahat}{\hat{A}}

\newcommand{\hsl}{\mathcal{H}_{sl}}

\newcommand{\rhosl}{\rho_{sl}}
\newcommand{\rhodl}{\rho_{dl}}
\newcommand{\msl}{\mathcal{M}^{sl}_x}
\newcommand{\mdl}{\mathcal{M}^{dl}_x}
\newcommand{\rsl}{\mathcal{R}^{sl}_x}
\newcommand{\rdl}{\mathcal{R}^{dl}_x}
\newcommand{\Gammah}{\Gamma ^{(h)}}
\newcommand{\tdsl}{T^{sl}_{(h)}}
\newcommand{\tddl}{T^{dl}_{(h)}}
\newcommand{\ie}{{\it i.e. }}
\newcommand{\cf}{{\it cf.~}}

\DeclareMathOperator{\End}{End}
\DeclareMathOperator{\Tr}{Tr}

\begin{document}

\title{Interplay of $SU(2)$, point group and translation symmetry for PEPS: application to a chiral spin liquid}

\author{Anna Hackenbroich}
\affiliation{Max-Planck-Institute of Quantum Optics, Hans-Kopfermann-Str. 1, D-85748 Garching, Germany}

\author{Antoine Sterdyniak}
\affiliation{Max-Planck-Institute of Quantum Optics, Hans-Kopfermann-Str. 1, D-85748 Garching, Germany}

\author{Norbert Schuch}
\affiliation{Max-Planck-Institute of Quantum Optics, Hans-Kopfermann-Str. 1, D-85748 Garching, Germany}

\begin{abstract}

Projected entangled pair states (PEPS) provide exact representations for many non-chiral topologically ordered states whereas their range of applicability to interacting chiral topological phases remains largely unsettled. In this context, the symmetries of the local PEPS tensors are crucial for determining the characteristic topological features of the state. In this article we examine the constraints that arise when different symmetries are imposed simultaneously on the local tensor such as internal $SU(2)$, point group and translation symmetry. We show how the interplay of these symmetries manifests in the entanglement spectrum which is the main diagnostic tool for chiral topological order. We apply our results to a spin liquid PEPS introduced previously as a chiral generalization of the resonating valence bond state. Our findings explain the discrepancies observed between the entanglement spectrum of this state and the expected edge spectrum described by a chiral conformal field theory. Finally, in a certain parameter region where this PEPS possesses an additional $U(1)$ symmetry we are able to resolve these discrepancies and obtain an entanglement spectrum with the expected state countings and conformal weight.

\end{abstract}
\date{\today}

\maketitle

\section{Introduction}

Topological phases are the most prominent examples of phases of matter that cannot be understood in terms of symmetry breaking and local order parameters.
In the strongly interacting regime, topological order can develop where the system hosts free anyonic bulk excitations~\cite{Laughlin1983,moore1991nonabelions}. While global symmetries may naively seem to play a smaller role for topological than for symmetry-broken phases, they are crucial for determining the different phases a system can realize as exemplified by the classification of free-fermionic~\cite{PhysRevB.78.195125,2009AIPC.1134...22K} and one-dimensional symmetry-protected~\cite{PhysRevB.83.075102,PhysRevB.84.165139,Chen1604} topological phases. Indeed, chiral topologically ordered phases~\cite{wen1990topological} possessing gapless edge modes described by chiral conformal field theory (CFT) can only appear in the absence of time reversal symmetry. They have been observed in the paradigmatic example of the fractional quantum Hall effect~\cite{tsui1982two,Laughlin1983} and were predicted to emerge in spin systems~\cite{kalmeyer1987,bauer2014chiral,gong2014emergent}. 

In the last decades, entanglement has emerged as a key probe for strongly correlated topological phases~\cite{kitaev2006topological,levin2006detecting,li2008entanglement}. The entanglement patterns characteristic for systems with local interactions at zero temperature~\cite{hastings2007area} are accurately captured by tensor network states~\cite{PhysRevLett.100.070502} such as matrix product states (MPS)~\cite{fannes1992finitely} in one dimension and projected entangled pair states (PEPS)~\cite{Verstraete:2004cf} in two dimensions. These variational states are defined in terms of local building blocks that mediate the entanglement between physical constituents through virtual particles. Crucially, they permit both analytical understanding~\cite{PhysRevB.84.165139} and numerically efficient algorithms~\cite{schollwock2011density,PhysRevB.94.035133,PhysRevB.97.045145}. In one dimension, variational algorithms approximating ground states with tensor network states are extremely successful even for critical systems~\cite{schollwock2011density} and algorithms based on PEPS have nowadays become competitive also in two dimensions~\cite{zheng2017stripe}. 

Despite their local structure, PEPS capture the physics of non-chiral topological order in a very simple and elegant manner. Indeed, ground states of many models with non-chiral topological order such as Kitaev's toric code~\cite{Kitaev2003} or string-net models~\cite{PhysRevB.71.045110} have exact representations in terms of simple PEPS~\cite{VerstraetePRL2006,gu2009tensor}. In these examples, topological order is encoded locally in the PEPS through symmetries of the virtual degrees of freedom~\cite{Schuch2010}. These virtual symmetries give direct access to characteristic properties of the topological phase such as the ground state manifold, topological entanglement entropy and fusion rules. 

In contrast, the application of the PEPS framework to chiral topological phases remains one of the open challenges in the field. It is known that Gaussian, \ie free fermionic, PEPS can possess a non-zero Chern number~\cite{wahl2013projected,Wahl2014PRB,DubailPRB2015}. However, they exhibit algebraically decaying correlation functions as was proven in a no-go theorem~\cite{DubailPRB2015,PhysRevB.95.115309}, implying that PEPS cannot exactly represent gapped chiral free-fermionic topological phases. As this no-go theorem does not apply to interacting PEPS, it is still unclear whether gapped chiral PEPS with intrinsic topological order exist. The two examples known to date are gapless: Firstly, in Ref.~\cite{Yang2015chiral}, a PEPS possessing the chiral CFT $\mathfrak{u}(1)_4$ as an edge theory was obtained by applying a Gutzwiller projection to two copies of a chiral Gaussian PEPS. Secondly, a chiral spin-liquid PEPS was constructed in Ref.~\cite{Poilblanc2015PRB_chiralPEPS,Poilblanc2016PRB_chiralPEPS} as a generalization of the square lattice resonating valence bond (RVB) state~\cite{anderson1987resonating,schuch2012resonating,Poilblanc2012PRB_RVB,PhysRevLett.111.037202} with long-range singlets and complex amplitudes. The entanglement spectrum (ES) of this PEPS resembles the spectrum of the chiral CFT $\mathfrak{su}(2)_1$ which is the edge spectrum of the bosonic Laughlin state at filling fraction $1/2$. However, certain discrepancies were observed between the PEPS entanglement spectrum and the CFT spectrum such as mismatching conformal weights and state-countings whose origin could not be resolved.

In the quest for PEPS representations of chiral topologically ordered states, the study of symmetries and their interplay is of particular importance: chiral spin liquids are invariant under multiple symmetries such as spin rotations, spatial rotations and translations. Moreover, their idealized instances transform equivalently under reflections and time reversal symmetry. For injective MPS and PEPS it is known that a given physical symmetry has to be represented locally on the virtual degrees of freedom~\cite{2006quant.ph..8197P,perez2010characterizing}. This understanding has led to crucial analytical results such as the classification of one-dimensional symmetry protected topological phases~\cite{PhysRevB.84.165139} as well as decisive speed-ups for numerical simulations~\cite{mcculloch2002non}. However, it has not been investigated systematically how multiple symmetries, for example spatial and internal transformations, can be implemented simultaneously and whether this leads to non-trivial constraints intrinsic to the PEPS formalism. 

In this article we analyze systematically the interplay of $\sutwo$, translation and point group symmetry for PEPS and then focus on the case of the chiral spin liquid PEPS which possesses an additional non-unitary symmetry, namely reflection combined with time-reversal. First, we show that for half-integer physical spins one cannot simultaneously impose invariance under spin-rotations, spatial rotations and single site translation at the level of the local tensor. The states obtained from local tensors satisfying either translation invariance or point group symmetry generally differ by their flux around non-contractible loops. Furthermore, in addition to the physical symmetries the local tensors necessarily possess a virtual $\mathbb{Z}_2$ symmetry which determines the possible topological properties of the state.

In the second part of this article, we consider the effect of the additional symmetry arising in the case of the chiral spin liquid PEPS. We show that its anti-unitarity has consequences in the transfer matrix and entanglement spectra, such as the emergence of a peculiar multiplet structure. These results explain the discrepancies of the ES from the CFT spectrum which were observed in Ref.~\cite{Poilblanc2015PRB_chiralPEPS,Poilblanc2016PRB_chiralPEPS}. In particular we show that the interplay of reflection symmetry and virtual $\mathbb{Z}_2$ symmetry leads to unphysical degeneracies in the PEPS entanglement spectrum. Finally, we show that there is a region of parameter space where the PEPS exhibits an additional virtual $\uone$ symmetry which permits to lift these degeneracies by considering states which break the underlying symmetry. We demonstrate numerically that within this region, the corrected low-energy ES of the PEPS is in perfect correspondence with the spectrum of the chiral CFT $\mathfrak{su}(2)_1$ including a correct value for the conformal weight.

The paper is organized as follows. In Sec.~\ref{sec:PEPSBasics} we introduce PEPS, recalling how global symmetries of the state such as space group and $\sutwo$ symmetries are implemented on the local tensors. We also review the role of virtual symmetries, the entanglement spectrum and the PEPS transfer matrix. In Sec.~\ref{sec:SymmCompatibility} we discuss the formal incompatibility of translation invariance and point group symmetry in $\sutwo$ invariant PEPS and analyze the consequences for half-integer spin. Using the example of the chiral PEPS from Ref.~\cite{Poilblanc2015PRB_chiralPEPS} we continue by analyzing the implications of $\sutwo$ invariance, translation invariance and point group symmetry for the transfer matrix and its fixed points in the case of a virtual $\mathbb{Z}_2$ symmetry in Sec.~\ref{sec:0Plus12PEPS}. Finally, in Sec.~\ref{sec:ZeroLambdaOne} we focus on the case where the PEPS possesses a virtual $\uone$ symmetry and present our numerical results establishing the correspondence between the CFT and entanglement spectra.

\section{Preliminaries on PEPS\label{sec:PEPSBasics}}

In this section we introduce our notations and review the construction of PEPS with space group symmetry, $\sutwo$ symmetry and virtual symmetries as well as the computation of entanglement spectra for cylinder PEPS.

\subsection{Construction of PEPS}

We study a spin system on a square lattice $\Lambda$ with one spin-$\frac{1}{2}$ degree-of-freedom per lattice site. The local Hilbert space on every lattice site is therefore two-dimensional and spanned by the states $\{\ket{\sx}|\sx=0,1\}$ with $\mathbf{x}\in\Lambda$. For every configuration $\{\mathbf{x}\mapsto\sx\}$ of the spins one obtains a many-body basis state $\ket{\{\mathbf{x}\mapsto\sx\}}$ for the whole system as the tensor product $\otimes_{\mathbf{x}\in\Lambda}\ket
{s_{\mathbf{x}}}$ of the corresponding local basis states on every lattice site. A generic quantum state
\begin{equation}
\ket{\psi}=\sum_{\{\mathbf{x}\mapsto\sx\}}c_{\{\mathbf{x}\mapsto\sx\}}\ket{\{\mathbf{x}\mapsto\sx\}}
\end{equation}
for the lattice spin system is defined by its expansion coefficients $c_{\{\mathbf{x}\mapsto\sx\}}$ with respect to this product basis. 

\begin{center}
 \begin{figure}[htb]
\begin{center}
\includegraphics[width= 0.99 \linewidth]{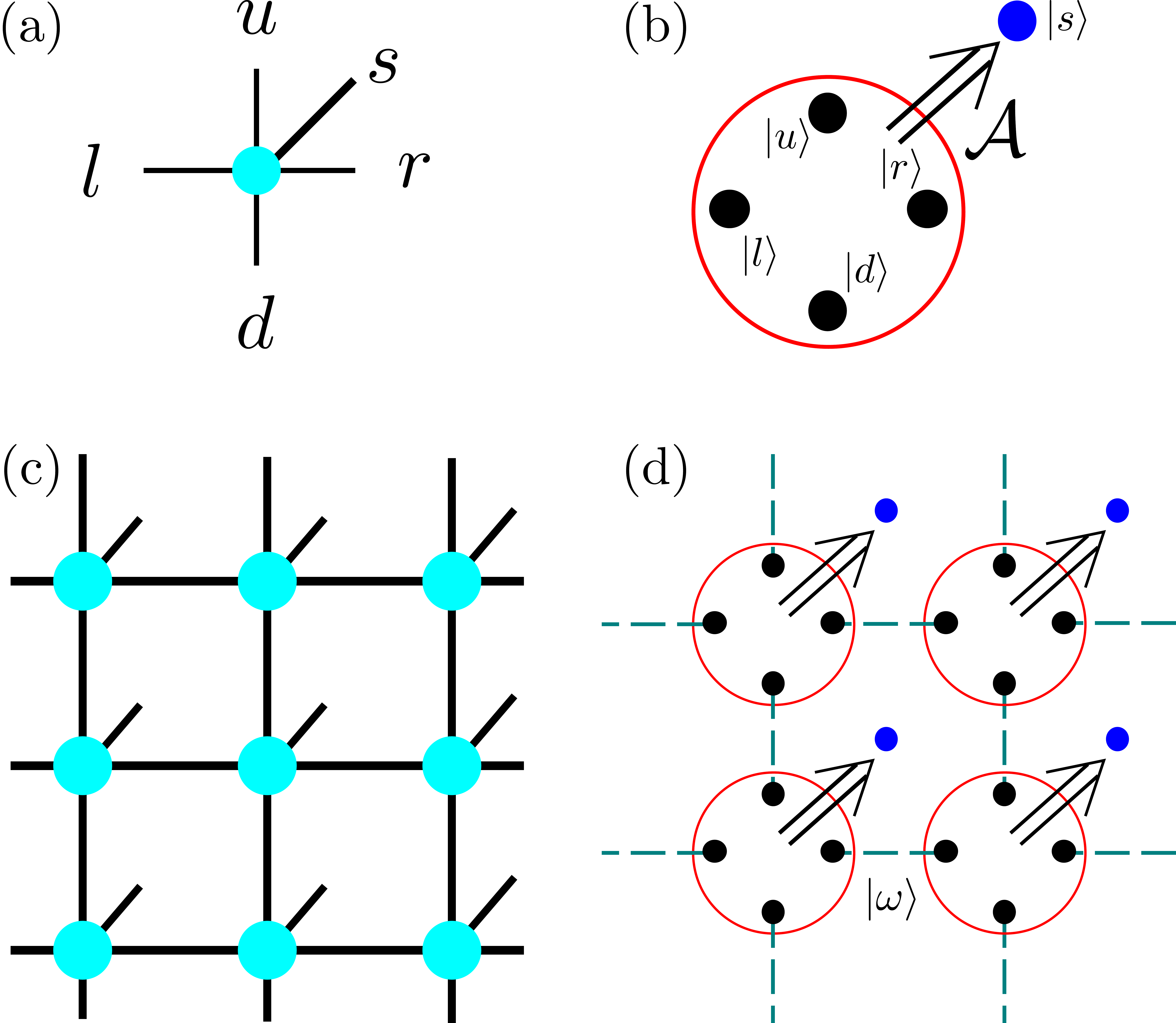}
\end{center}
\caption{Schematic description of the two different ways of defining a PEPS. (a): Four virtual spins are introduced around each physical spin. The relation between them is given by a five-index tensor. The many-body state is obtained by contracting nearest-neighbour virtual spins as shown in (c). (b): The PEPS is defined by a local projection map from the four virtual spins to the physical spin. Nearest-neighbour virtual spins are placed in a maximally entangled state $\ket{\omega}$ as shown in (d).}
\label{fig::PEPS_definition} 
\end{figure}
\end{center}

Projected entangled pair states (PEPS) are model states for lattice spin systems which depend only on a small number of parameters. These are given by the entries $A^s_{lurd}$ of a five-index tensor that describes the physical spin $s$ of one lattice site as well as four virtual spins $l,u,r,d$ placed at the left, top, right and bottom of each lattice site, respectively (see Fig.~\ref{fig::PEPS_definition}(a)). The dimension $D$ of the Hilbert space for each virtual spin is called the bond dimension and is independent of the dimension $d=2$ of the physical spin-$\frac{1}{2}$ Hilbert space on every site. The network obtained by placing a tensor $A(\mathbf{x})$ on each lattice site $\mathbf{x}$ and contracting nearest-neighbour virtual indices defines the PEPS expansion coefficient
\begin{equation}\label{PEPSBasisCoefficient}
c_{\{\mathbf{x}\mapsto\sx\}} =  \sum_{\substack{\{l_{\mathbf{x}}u_{\mathbf{x}}\\r_{\mathbf{x}}d_{\mathbf{x}}\}}}\prod_{\mathbf{x}\in\Lambda} \delta_{l_{\mathbf{x}},r_{\mathbf{x}-\mathbf{e}_1}}\delta_{u_{\mathbf{x}},d_{\mathbf{x}+\mathbf{e}_2}} A(\mathbf{x})^{\sx}_{l_{\mathbf{x}}u_{\mathbf{x}}r_{\mathbf{x}}d_{\mathbf{x}}}
\end{equation}
where $l_{\mathbf{x}},u_{\mathbf{x}},r_{\mathbf{x}},d_{\mathbf{x}}=0,\dotsc,D-1$ for all lattice sites $\mathbf{x}\in\Lambda$ (see Fig.~\ref{fig::PEPS_definition}(c)). In Eq.~\eqref{PEPSBasisCoefficient} we implicitly assumed periodic boundary conditions $r_{\mathbf{x}}=l_{\mathbf{x}+(N_h-1)\mathbf{e}_1}$ ($d_{\mathbf{x}}=u_{\mathbf{x}+(N_v-1)\mathbf{e}_2}$) for the virtual spins corresponding to a torus of size $N_h$ ($N_v$) in the horizontal (vertical) direction. A PEPS defined on a torus therefore has no open virtual legs and is a state of the physical spins only. In the following we also study PEPS on cylinders obtained by compactifying only the vertical direction. The choice of virtual boundary conditions at the left and right edges of the cylinder may have a profound impact on the resulting physical state which persists even in the limit of an infinitely long cylinder $N_h\rightarrow\infty$. For instance whenever the local tensor possesses some symmetry the quantum numbers of the virtual boundary vectors influence the transformation behaviour of the physical state. The degrees of freedom for a PEPS on the cylinder therefore consist of the physical spins and the boundary virtual spins.

Alternatively, PEPS can be defined as the result of a projection of a layer of entangled virtual spins onto the layer of physical spins. As above one associates to every lattice site four $D$-dimensional virtual spins placed at its left, top, right and bottom edges which are then mapped to the physical spin of this site by the local tensor map (see Fig.~\ref{fig::PEPS_definition}(b))
\begin{equation}\label{localprojectionmap}
\amap=\sum_{s}\sum_{l,u,r,d=0}^{D-1} \ahat^s_{lurd} \ket{s}\big[\bra{l}\otimes\bra{u}\otimes\bra{r}\otimes\bra{d}\big].
\end{equation}
Throughout this paper we also refer to $\amap$ as a local projection map and we denote its basis entries by $\ahat$ in order to distinguish them from the local tensor $A$. Here, $\{\ket{l}| l =0, \dotsc, D-1\}$ is an orthonormal basis of the Hilbert space of the left virtual spin and similarly for $\ket{u}$ (up), $\ket{r}$ (right) and $\ket{d}$ (down). Nearest-neighbour virtual spins on adjacent lattice sites $\mathbf{x},\mathbf{y}$ are placed in a pairwise maximally entangled state $\ket{\omega(\mathbf{x},\mathbf{y})} = \sum_{i_{\mathbf{x}},j_{\mathbf{y}} = 0}^{D-1} \omega_{i_{\mathbf{x}}j_{\mathbf{y}}}\ket{i_{\mathbf{x}}}\otimes\ket{j_{\mathbf{y}}}$. As sketched in Fig.~\ref{fig::PEPS_definition}(d) the application of the product of all local projection maps to the tensor product of the virtual maximally entangled states for all nearest-neighbour bonds $\langle\mathbf{x},\mathbf{y}\rangle$ defines the PEPS
\begin{equation}
\ket{\tilde{\psi}} = \big[\bigotimes_{\mathbf{x}\in \Lambda}\amap(\mathbf{x})\big] \prod_{\langle\mathbf{x},\mathbf{y}\rangle}\ket{\omega(\mathbf{x},\mathbf{y})}
\end{equation}with many-body basis coefficients
\begin{equation}\label{PEPSBasisCoefficient2}
\tilde{c}_{\{\mathbf{x}\mapsto\sx\}} =  \sum_{\substack{\{l_{\mathbf{x}}u_{\mathbf{x}}\\r_{\mathbf{x}}d_{\mathbf{x}}\}}}\prod_{\mathbf{x}\in\Lambda} \omega_{l_{\mathbf{x}},r_{\mathbf{x}-\mathbf{e}_1}}\omega_{u_{\mathbf{x}},d_{\mathbf{x}+\mathbf{e}_2}} \ahat(\mathbf{x})^{\sx}_{l_{\mathbf{x}}u_{\mathbf{x}}r_{\mathbf{x}}d_{\mathbf{x}}}.
\end{equation}
If the basis entries of the tensor map coincide with the local tensor, \ie $\ahat^s_{lurd} = A^s_{lurd}$, and all virtual maximally entangled states are given by $\ket{\omega}=\sum_{i=0}^{D-1}\ket{i}\otimes\ket{i}$ with basis entries $\omega_{ij}=\delta_{ij}$ the many-body basis coefficients~\eqref{PEPSBasisCoefficient} and \eqref{PEPSBasisCoefficient2} agree such that the states $\ket{\psi}$ and $\ket{\tilde{\psi}}$ are identical. However, other options for $\ahat$ and $\ket{\omega}$ exist and are relevant for the construction of $\sutwo$ spin-singlet PEPS. The PEPS in Eq.~\eqref{PEPSBasisCoefficient2} can be cast into the form~\eqref{PEPSBasisCoefficient} with local tensors $A$ defined by absorbing the virtual maximally entangled state $\omega_{ij}$ into $\ahat$. For the rest of this section we therefore assume w.l.o.g. that the maximally entangled state is given by $\omega_{ij}=\delta_{ij}$ and $\ahat^s_{lurd} = A^s_{lurd}$ unless stated otherwise.

\subsection{PEPS with space group symmetry\label{sec:SpaceGroup}}

In this subsection we examine how lattice translations and point group transformations act on PEPS. All lattice rotations and reflections considered here are defined with respect to the vertices of the lattice. An element $\ket{\{\mathbf{x}\mapsto\sx\}}=\otimes_{\mathbf{x}\in\Lambda}\ket{s_{\mathbf{x}}}$ of the many-body product basis of a lattice spin system is mapped by a space group transformation $g$ to a different basis state $\ket{\{\mathbf{x}\mapsto s_{g^{-1}\mathbf{x}}\}}=\otimes_{\mathbf{x}\in\Lambda}\ket{s_{g^{-1}\mathbf{x}}}$. Such operations therefore map any PEPS to a different PEPS whose expansion coefficients are obtained from transformed local tensors
\begin{equation}\label{TrafoLocalTensor}
\tilde{A}(\mathbf{x})^{\sx}_{lurd}=A(g^{-1}\mathbf{x})^{\sx}_{g(lurd)}.
\end{equation}
Here, the space group action on the virtual indices is trivial for translations whereas for the point group it is given by the natural two-dimensional representation of $C_{4v}$ on the directions left, up, right and down. In the following we consider only translation-invariant PEPS but with unit cells which can be larger than that of the underlying spin lattice $\Lambda$.
 
We recall that the point group $C_{4v}$ for a site $\mathbf{x}\in\Lambda$ of the square lattice has four real one-dimensional representations denoted by boldface letters $\mathbf{A}_1,\mathbf{A}_2,\mathbf{B}_1,\mathbf{B}_2$. In the following we will be especially interested in the rotation-invariant representations $\mathbf{A}_1,\mathbf{A}_2$ that are even and odd under mirrors, respectively. Based on Eq.~\eqref{TrafoLocalTensor} we say that a PEPS local tensor $A$ transforms in a one-dimensional representation $\sigma$ of $C_{4v}$ if
\begin{equation}\label{PointGroupLocalTensor}
A^s_{g(lurd)}=\sigma(g)A^s_{lurd}
\end{equation}
for all point group elements $g$. 

Whenever the local tensor of a translation-invariant PEPS is point-group symmetric according to Eq.~\eqref{PointGroupLocalTensor}, the local representation $\sigma$ also determines the transformation behaviour of the PEPS under $C_{4v}$ transformations. We denote the resulting point group representation of the PEPS by $\Sigma(\sigma)$ which generally depends also on the system size. For instance if the local tensor transforms as $\sigma=\mathbf{A}_1$ the resulting PEPS will also be invariant under rotations and reflections, \ie $\Sigma (\mathbf{A}_1)=\mathbf{A}_1$. On the other hand, a local representation $\sigma =\mathbf{A}_2$ causes the PEPS to transform under $C_{4v}$ in the representation $\Sigma(\mathbf{A}_2)=\mathbf{A}_1$ ($\Sigma(\mathbf{A}_2)=\mathbf{A}_2$) on a lattice with an even (odd) number of sites. These statements extend to PEPS with a checkerboard sub-lattice structure which are translation invariant with a bigger unit cell of $2\times 2$ lattice sites. The checkerboard lattice is invariant under $C_{4v}$ operations. If both local tensors of such a PEPS satisfy Eq.~\eqref{PointGroupLocalTensor} the state will therefore transform in the same representation $\Sigma(\sigma)$ as a translation-invariant PEPS whose local tensor has the representation $\sigma$. In addition to Eq.~\eqref{PointGroupLocalTensor} there generally are other possibilities to ensure that a PEPS transforms under the point group in a representation $\Sigma$; for instance it suffices that the local tensor transforms in the representation $\sigma$ up to a local basis change of the virtual spins.  

In the final sections of this article, we study a chiral spin-liquid PEPS which is invariant under lattice translations and rotations but gets mapped to its complex conjugate by lattice mirrors. Its real (imaginary) part therefore transforms under $C_{4v}$ in the one-dimensional representation $\mathbf{A}_1$ ($\mathbf{A}_2$) and we denote the transformation of the entire state by $\mathbf{A}_1+i\mathbf{A}_2$. Such a state is special since the transformation behaviour under time reversal and lattice symmetries which is expected for the edge modes of a chiral topological system is satisfied by the bulk of the PEPS. By analogy we say that a PEPS local tensor 
\begin{equation}\label{A1+iA2LocalTensor}
A^s_{lurd} = (A_1)^s_{lurd} + i\,(A_2)^s_{lurd}
\end{equation}
transforms under the point group as $\sigma =\mathbf{A}_1+i\mathbf{A}_2$ if the real tensors $A_1$ and $A_2$ transform in the representations $\mathbf{A}_1$ and $\mathbf{A}_2$ according to Eq.~\eqref{PointGroupLocalTensor}, respectively. The local tensor $A$ is then invariant (complex conjugated) under cyclic permutations (reflections) of its virtual indices. One can show that a translation invariant PEPS whose local tensor satisfies Eq.~\eqref{A1+iA2LocalTensor} transforms under $C_{4v}$ in the representation $\Sigma (\mathbf{A}_1+i\mathbf{A}_2)=\mathbf{A}_1+i\mathbf{A}_2$ regardless of the system size.  

\subsection{Spin-singlet PEPS\label{sec:SpinSingletPEPS}}  

Within the PEPS framework it is possible to construct quantum states which are invariant under global group transformations such as $\sutwo$. This is the case if two conditions are met~\cite{perez2010characterizing,2018arXiv180404964M}. Firstly, the local projection map~\eqref{localprojectionmap} should be an intertwiner between the group representation $\rho_{phys}$ of the physical spin and some group representations $\rho_{v,l}$ $(\rho_{v,u},\rho_{v,r},\rho_{v,d})$ for the left (up, right, down) virtual spins such that
\begin{equation}\label{intertwiner}
\rho_{phys}(\gamma)\circ \amap = \amap \circ \big[\rho_{v,l}(\gamma)\otimes\rho_{v,u}(\gamma)\otimes\rho_{v,r}(\gamma)\otimes\rho_{v,d}(\gamma)\big]
\end{equation}
for all group elements $\gamma$. This implies that any group operation acting on the physical spins can be pushed to the virtual layer where it factorises as a product of group operations acting on every virtual leg separately. Secondly, if a virtual particle transforms in the representation $\rho_v$ its nearest-neighbour virtual spin has to transform in the conjugate virtual representation $\rho_v^*$ with basis representation $\rho_v^*(\gamma)_{ij}=\rho_v(\gamma^{-1})_{ji}$ for any group element $\gamma$. The contribution of this nearest-neighbour virtual bond to the many-body basis coefficient~\eqref{PEPSBasisCoefficient} is thus invariant under group transformations since $\rho_v(\gamma)_{i'i}\delta_{ij}\rho_v^*(\gamma)_{j'j}=(\rho_v(\gamma)\circ\rho_v(\gamma^{-1}))_{i'j'}=\delta_{i'j'}$. The basis coefficient remains therefore unchanged and the PEPS transforms trivially under global group operations.  

Generally, invariance of the PEPS therefore prevents nearest-neighbor virtual spins such as the left and right (up and down) tensor legs from transforming in identical group representations. However, one can reformulate a pair of nearest-neighbour virtual spins with representations $(\rho_v,\rho_v^*)$ as two identical representations $(\rho_v,\rho_v)$ by simultaneously changing their connecting maximally entangled state provided that $\rho_v$ is self-conjugate. Let us exemplify this reformulation for the Lie group $\sutwo$ whose representations are self-conjugate with the isomorphism between a representation and its conjugate given by the spin-flip operator $Y$,
\begin{equation}\label{SelfConjugacy}
\rho^*(\gamma) = Y\circ \rho(\gamma) \circ Y^{-1}.
\end{equation}
We consider the horizontal bond between two sites $\mathbf{x}$ and $\mathbf{x'}=\mathbf{x}+\mathbf{e}_1$ with local tensors $\hat{A}(\mathbf{x})$ and $\hat{A}(\mathbf{x'})$ initially connected with an identity on the virtual bond. $\sutwo$ invariance then requires that the right (left) virtual spin of site $\mathbf{x}$ ($\mathbf{x'}$) transforms in the representation $\rho_v$ ($\rho_v^*$). Without changing the PEPS, we can insert an identity $\mathbbm{1} = Y Y^{-1}$ into the virtual bond and define a modified right local tensor by absorbing the inverse spin flip to its left,
\begin{equation}\label{SingletAbsorptionOne}
\tilde{\hat{A}}(\mathbf{x'})^{s'}_{l'u'r'd'}= \sum_{\tilde{l}'}(Y^{-1})_{l'\tilde{l}'}\hat{A}(\mathbf{x'})^{s'}_{\tilde{l}'u'r'd'}.
\end{equation}
Due to the self-conjugacy Eq.~\eqref{SelfConjugacy} the left virtual leg of $\tilde{\hat{A}}$ transforms in the same representation $\rho_v$ as its nearest neighbour. However, the two local tensors are now contracted with a non-trivial virtual maximally entangled state $\omega_{ij}=Y_{ij}$ that we also refer to as a virtual singlet,
\begin{equation}
\hat{A}(\mathbf{x})
^s_{lurd} \delta_{rl'} \hat{A}(\mathbf{x'})^{s'}_{l'u'r'd'} = \hat{A}(\mathbf{x})
^s_{lurd} \omega_{rl'} \tilde{\hat{A}}(\mathbf{x'})^{s'}_{l'u'r'd'}.
\end{equation}

This reformulation is advantageous for the systematic study of PEPS with simultaneous $\sutwo$ and space group symmetry. Indeed, it allows the construction of $\sutwo$ invariant PEPS in terms of local projection maps whose four virtual spins transform identically under spin rotations. This is the situation we will study in the following, \ie we place virtual singlets on all bonds and moreover assume that the horizontal and vertical representations are identical. We can then investigate the symmetry properties of the local projection map and those of the virtual singlets separately. In the canonical $S^z$ eigenbasis the spin-flip operator is given by the unitary
\begin{equation}
Y = e^{i\pi S^y}.
\end{equation}
It is symmetric (antisymmetric) and squares to $+1$ ($-1$) for integer (half-integer) spin representations. The virtual singlet $\ket{\omega}$ is therefore neither symmetric nor anti-symmetric under exchange of the two virtual particles if $\rho_v$ contains a mixture of integer and half-integer spin representations. As will be discussed in Sec.~\ref{sec:VirtualSymmetry}, this is a necessary condition for the realization of PEPS with half-integer physical spin. In this case there may be several distinct spin-singlet PEPS with the same local projection map and inequivalent orientations for the virtual singlets. We devote Sec.~\ref{sec:SymmCompatibility} to the study of their relations and symmetry properties. We emphasize that the singlet absorption Eq.~\eqref{SingletAbsorptionOne} relies on the self-conjugacy of $\rho_v$ and is therefore not always possible for groups possessing non-self-conjugate representations such as SU(3).

\subsection{Virtual symmetries of PEPS\label{sec:VirtualSymmetry}}

Many toy models with intrinsic topological order such as the $\mathbb{Z}_2$ toric code~\cite{Kitaev2003} possess exact representations in terms of simple PEPS~\cite{VerstraetePRL2006}. For these models it has been realized that topological order is intimately related to the invariance of the local tensor under virtual symmetries which by definition do not involve the physical spin~\cite{Schuch2010},
\begin{equation}
\amap = \amap \circ \big[\tau_{v,l}(g)\otimes \tau_{v,u}(g)\otimes \tau_{v,r}(g)\otimes \tau_{v,d}(g)\big] \label{eq:virtual_sym}\textrm{.}
\end{equation} 
Here, $\tau_{v,l}$ $(\tau_{v,u},\tau_{v,r},\tau_{v,d})$ are representations of the virtual symmetry group $G$ carried by the left (up, right, down) virtual legs with $\tau_{v,l}^*=\tau_{v,r}$ and $\tau_{v,d}^*=\tau_{v,u}$. For such PEPS all characteristic features of intrinsic topological order such as a topological ground state degeneracy, topological entanglement entropy or anyonic excitations can be traced back to 
the virtual symmetry~\eqref{eq:virtual_sym} of the local tensor. We review in particular the construction of states from the topological ground state manifold that are locally equivalent to a PEPS $\ket{\psi}$ but generally possess different eigenvalues with respect to certain non-local operators.

\begin{center}
 \begin{figure}[htb]
\begin{center}
\includegraphics[width= 0.95 \linewidth]{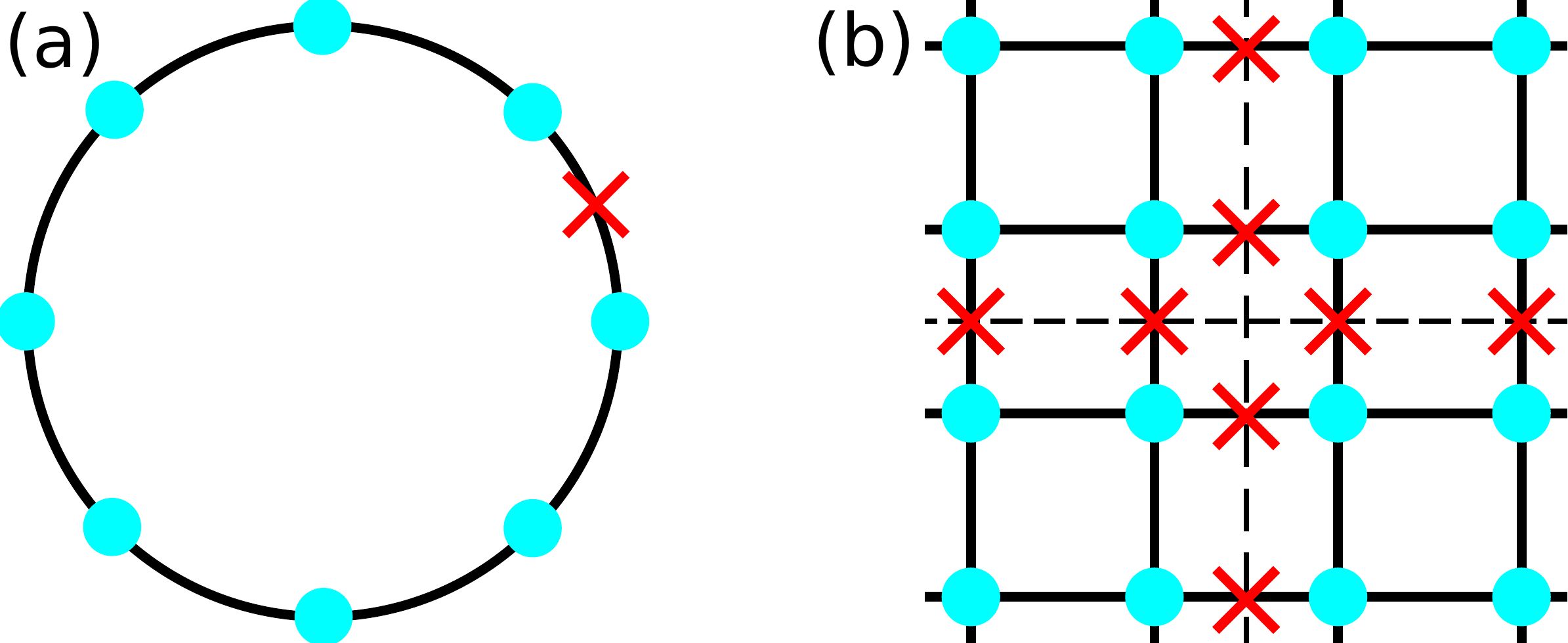}
\end{center}
\caption{Insertion of $\mathbb{Z}_2$ fluxes into one- and two-dimensional tensor network states by multiplication of certain virtual bonds with matrices $Z$ (here represented by red crosses). The blue points represent the local tensors for which we suppressed the physical legs in the interest of readability. (a) MPS $\ket{\psi}_Z$ with a non-trivial flux through the circle. (b) PEPS $\ket{\psi}_{h,v}$ with both horizontal and vertical flux lines. Similarly there exist PEPS $\ket{\psi}_{h}$, $\ket{\psi}_{v}$ with only one horizontal or vertical flux line, respectively.}
\label{fig::MPS_PEPS_flux} 
\end{figure}
\end{center}

On a square lattice torus or cylinder, vertical (horizontal) flux lines of the virtual symmetry group $G$ can be added to the PEPS $\ket{\psi}$ by inserting a group element $\tau_{v,r(u)}(g)$ on every horizontal (vertical) link crossed by a vertical (horizontal) line through the centres of plaquettes as sketched in Fig.~\ref{fig::MPS_PEPS_flux}(b). Due to the virtual symmetry~\eqref{eq:virtual_sym} of every local tensor, these strings can be moved throughout the bulk of the tensor network. Hence they are not localized at any one position and generally cannot be detected by a local operator such as a local Hamiltonian. The number of different states that can be generated through addition of flux strings depends both on the group $G$ and on the topology of the underlying lattice~\cite{PhysRevB.41.9377,PhysRevB.71.045110,Kitaev2003}. Indeed, the former determines the number of string types whereas the latter determines the number of independent non-contractible loops. However this consideration gives only the maximal dimension of the ground state manifold as some states may vanish or be linearly dependent. This notably occurs at phase transitions~\cite{Schuch2013PRL}.

Let us now focus on $\sutwo$ invariant PEPS with half-integer physical spin per unit cell which necessarily possess a virtual symmetry with symmetry group $G=\mathbb{Z}_2$. Indeed, the tensor product of two integer or two half-integer spin representations contains only integer spins. As a result, the intertwiner condition~\eqref{intertwiner} between physical and virtual $\sutwo$ representations has a solution for half-integer physical spin only if the virtual representation $\rho_v$ contains both integer and half-integer spins. The $\sutwo$ rotation $e^{2\pi i S^z}$ then has a non-trivial virtual representation in terms of a diagonal matrix
\begin{equation}
Z = \rho_v (e^{2\pi i S^z})
\end{equation} 
equal to $+1$ ($-1$) on integer (half-integer) virtual spin representations such that $Z\neq \pm \mathbbm{1}$ but $Z^2 = \mathbbm{1}$. Due to its $\sutwo$ invariance the local projection map for half-integer physical spin satisfies
\begin{equation}\label{VirtualZ2Symmetry}
(-1)\times \amap = \amap \circ Z^{\otimes 4}
\end{equation} 
in analogy to Eq.~\eqref{eq:virtual_sym}. The overall sign $-1$ stems from the half-integer physical spin and has no essential influence on the intrinsic topological features of the resulting PEPS.

For PEPS with a virtual $\mathbb{Z}_2$ symmetry there is only a single type of flux string that can be inserted around any non-contractible loop since two strings around the same loop annihilate each other. This leads to an expected topological degeneracy of four in a system defined on a two-dimensional torus, where the ground state manifold is spanned by the original state $\ket{\psi}$ as well as $\ket{\psi}_{h}$, $\ket{\psi}_{v}$ and $\ket{\psi}_{h,v}$ with horizontal, vertical and both horizontal and vertical flux lines, respectively. Similar concepts apply to matrix product states (MPS) as the one-dimensional analogues of PEPS. If the local MPS tensor possesses a virtual $\mathbb{Z}_2$ symmetry one can define a state $\ket{\psi}_Z$ with a non-vanishing flux through the circle by inserting a matrix $Z$ on one virtual bond as sketched in Fig.~\ref{fig::MPS_PEPS_flux}(a). 

\subsection{Entanglement spectrum for PEPS\label{sec:ESPEPS}}

Ever since their introduction~\cite{li2008entanglement}, entanglement spectra have been used extensively to probe the nature of states especially in the context of topologically ordered phases~\cite{sterdyniak2011extracting,sterdyniak2013series,PhysRevX.1.021014}. The entanglement spectrum (ES) of a part $A$ of a system in the state $\ket{\psi}$ is defined as the spectrum of its entanglement Hamiltonian $H_{Ent}=-\log \rho_A$ where $\rho_A = \Tr_{\bar{A}} \ket{\psi}\bra{\psi}$ is the corresponding reduced density matrix. In the rest of this paper we will focus on the case where $A$ and $\overline{A}$ form a real-space bi-partition of the system. The ES is related to the spectrum of the physical edge theory as shown numerically in many cases~\cite{thomale2010entanglement,sterdyniak2012real,PhysRevB.85.115321,PhysRevLett.115.116803} and analytically for non-interacting topological phases~\cite{PhysRevLett.104.130502}, certain one-dimensional symmetry protected topological phases~\cite{PhysRevB.81.064439} and for quantum states whose edge states are described by a chiral conformal field theory~\cite{PhysRevLett.108.196402}. This allows for example to extract conformal data such as conformal weights and thus to identify the edge conformal theory starting only from the ground state~\cite{lauchli2013operator}.

\begin{center}
 \begin{figure}[htb]
\begin{center}
\includegraphics[width= 0.8 \linewidth]{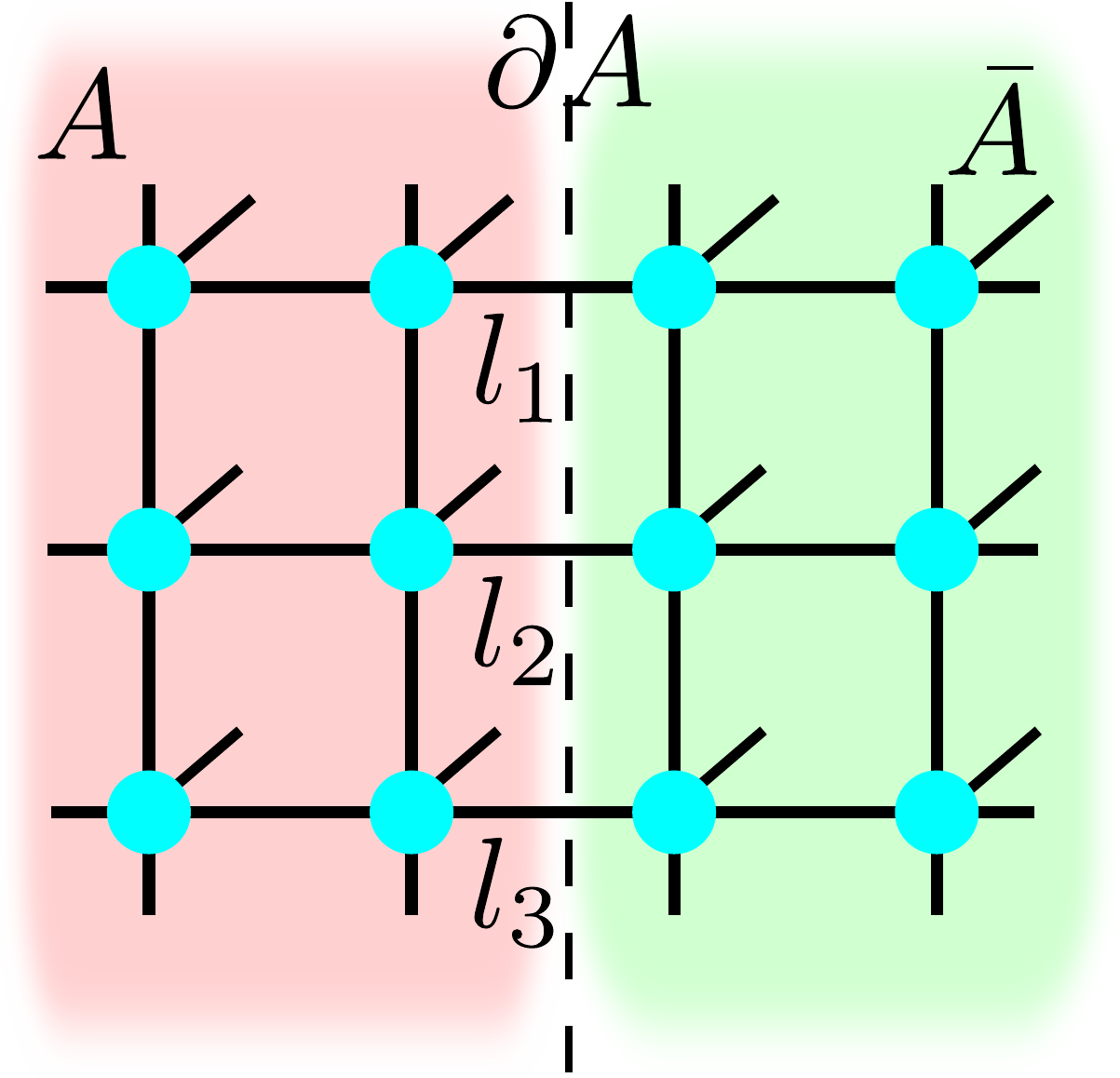}
\end{center}
\caption{Entanglement spectrum for PEPS. The lattice is divided in two disjoint regions $A$ and $\bar{A}$. The entanglement between these regions is naturally carried by the virtual spins at the one-dimensional boundary. The entanglement Hamiltonian is obtained as an operator acting only on those spins (Eq.~\ref{eq:ent:Ham}).}
\label{fig::PEPS_ES} 
\end{figure}
\end{center}

A correspondence at the level of the ES between bulk and edge degrees of freedom is very natural in the framework of PEPS. We can decompose the state for the entire system as a sum
\begin{equation}\label{EntanglementDecomposition}
\ket{\psi} = \sum _{\mathbf{l}} \ket{\psi(\mathbf{l})}_A \otimes \ket{\psi(\mathbf{l})}_{\bar{A}} 
\end{equation} 
where $\ket{\psi(\mathbf{l})}_{A(\bar{A})}$ denotes the PEPS restricted to the subsystem $A$ ($\bar{A}$) depending explicitly on the configuration $\mathbf{l}=(l_1,l_2,l_3,\dotsc)$ of the virtual legs which cross the entanglement cut $\partial A$ (see Fig.~\ref{fig::PEPS_ES}). Eq.~\eqref{EntanglementDecomposition} shows that the entanglement between the physical spins in $A$ and $\bar{A}$ is carried by the virtual spins connecting the two regions. Within the PEPS formalism it is therefore natural to interpret the entanglement Hamiltonian of a two-dimensional system as an operator for the virtual spins at the one-dimensional boundary~\cite{cirac2011entanglement}.

Specifically, we consider a PEPS on a cylinder of total length $N_h$ with sub-system $A$ consisting of the first $N_A$ columns starting from the left edge. The entanglement cut therefore crosses all $N_v$ horizontal virtual bonds of one PEPS column with Hilbert space $\hsl= (\mathbb{C}^D)^{\otimes N_v}$ where the subscript indicates that all virtual bonds lie in a single layer. We stack the PEPS with its complex conjugate, thereby forming an object with both a ket-layer and a bra-layer, and compute the reduced density matrix $\rho_A$ by contracting the physical legs corresponding to sites in the complement $\bar{A}$. A detailed calculation shows that~\cite{cirac2011entanglement}
\begin{equation}\label{eq:ent:Ham}
 \rho_A = U \sqrt{(\sigma^L)^T} \sigma^R \sqrt{(\sigma^L)^T} U^{\dagger}
\end{equation}
where $U$ is an isometry from the boundary virtual spins $\hsl$ to the physical spins in $A$. On the other hand, the virtual reduced density matrix $\sigma^{L(R)}$ is an operator which maps the virtual spins at the left (right) edge of the entanglement cut from the bra-layer to the ket-layer. It is obtained by tracing out the physical spins in the density matrix of the restricted state $\ket{\psi(\mathbf{l})}_A$ ($\ket{\psi(\mathbf{l})}_{\bar{A}}$) while keeping the virtual legs at the entanglement cut free, 
\begin{equation}
\sigma^{L}_{\mathbf{l}\mathbf{\tilde{l}}} = \Tr_{A} \big[\ket{\psi(\mathbf{l})}^{}_{A} \prescript{}{A}{\langle \psi(\mathbf{\tilde{l}}) |}\big]
\end{equation}
and similarly for $\sigma^R$. Here, all virtual reduced density matrices are normalized according to $\Tr (\sigma^{L(R)})^2 = 1$. In this paper we study reflection-symmetric PEPS for which the left and right virtual density matrices are related as $(\sigma^R)^*= \sigma^L$ such that the ES is given by the spectrum of
\begin{equation}\label{ESHermitianDensityMatrix}
-\log \big[(\sigma^L)^2\big].
\end{equation}  

\subsection{Transfer matrix\label{sec:TransferMatrix}}

\begin{center}
 \begin{figure}[htb]
\begin{center}
\includegraphics[width= 0.7 \linewidth]{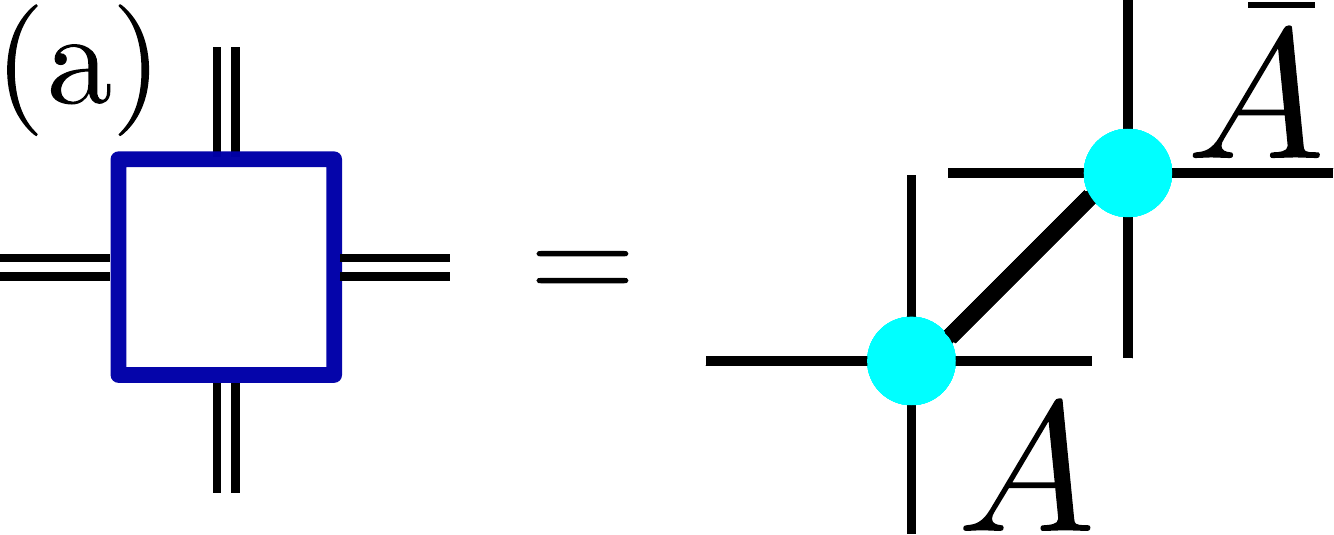}\\
\includegraphics[width= 0.2 \linewidth]{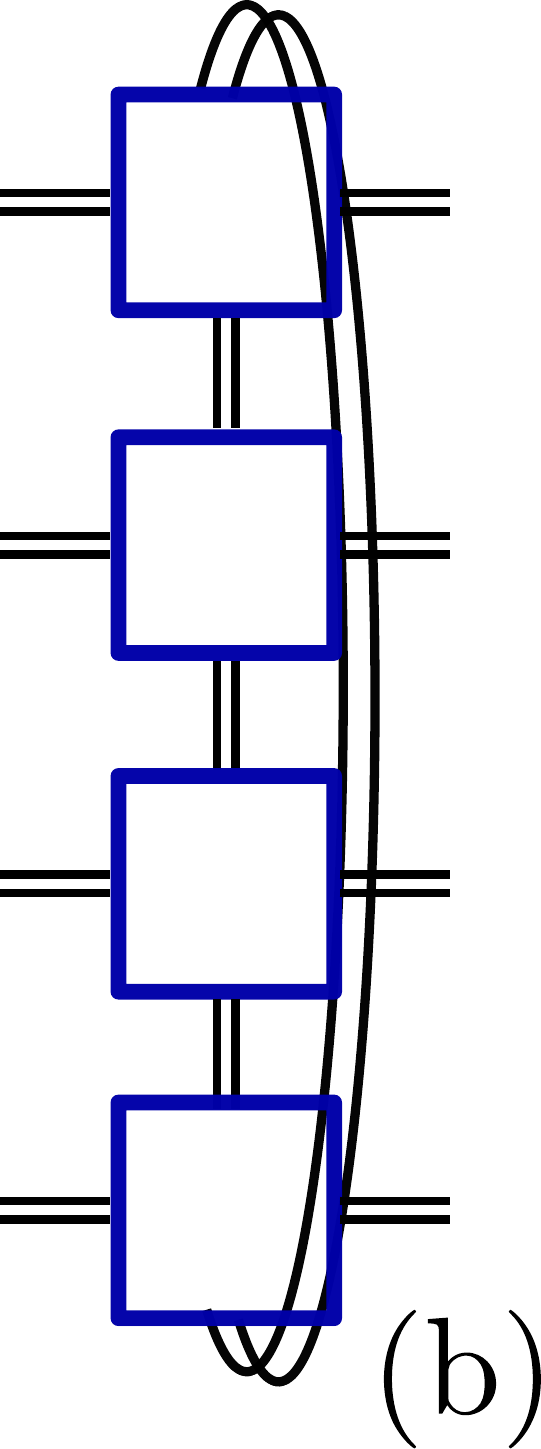}
\end{center}
\caption{Definition of the cylinder transfer matrix. (a) The local tensor is contracted with its complex conjugate to obtain the single-site transfer matrix. (b) Single-site transfer matrices are contracted along the periodic direction of the cylinder to obtain the cylinder transfer matrix.}
\label{fig::transfer_matrix} 
\end{figure}
\end{center}

An object of central importance for the study of PEPS in two dimensions is the cylinder transfer matrix $\Gamma$. As sketched in Fig.~\ref{fig::transfer_matrix}, $\Gamma$ is obtained by stacking a single column of local PEPS tensors with their complex conjugates and contracting the physical indices of both layers. As for MPS, the transfer matrix spectrum determines the correlation length of the state. Moreover, in the presence of virtual symmetries the leading eigenvalues of $\Gamma$ in different symmetry sectors determine the number of independent ground states with non-vanishing norm~\cite{Schuch2013PRL}. We will make use of this fact in Sec.~\ref{sec:SpectrumTransferMatrixUOne}.

In addition the transfer matrix gives access to the virtual reduced density matrices and thereby to the ES of the PEPS. If the state has virtual boundary conditions $v^{L(R)}\in\hsl$ at the left (right) edges of the cylinder one finds
\begin{subequations}\label{VirtualDensityMatrixTransferMatrix}
\begin{gather}
\sigma^L = (\Gamma^T)^{N_A}\big(v^L\otimes (v^L)^*\big),\\
\sigma^R = \Gamma^{N_h-N_A}\big(v^R\otimes (v^R)^*\big).
\end{gather}
\end{subequations}
Here we used that the transfer matrix acts equivalently on two copies $\hsl\otimes\hsl$ of the single-layer virtual column or on virtual density matrices $\sigma\in\End(\hsl)$. For an infinitely long cylinder the subleading eigenspaces of the transfer matrix are suppressed in Eq.~\eqref{VirtualDensityMatrixTransferMatrix} and $\sigma^{L,R}$ correspond to the leading left- and right-eigenvectors. 

Without symmetries or fine-tuning, these leading eigenvectors are non-degenerate, positive and have non-vanishing overlap with a generic boundary vector. The ES is therefore expected to be independent of the boundary conditions in the thermodynamic limit. However, in the presence of virtual symmetries the transfer matrix is block diagonal and the ES may depend on the symmetry sector of the boundary vectors. All in all, the transfer matrix therefore contains crucial information about the topological properties of a PEPS and we devote Sec.~\ref{sec:SymmetriesTransferMatrix} to the study of its symmetries for the chiral PEPS in which we are interested.

\section{Incompatibility of translation invariance and point group symmetry in construction of $\sutwo$ invariant PEPS\label{sec:SymmCompatibility}}

Spin liquids are states which are invariant under global $\sutwo$ transformations and lattice translations in addition to transforming in a well-defined way under rotations and reflections about lattice sites. In order to construct spin liquid PEPS we therefore need to understand how translation invariance and point group symmetry can be implemented for $\sutwo$ invariant PEPS. This is a subtle issue since generically either translation invariance or point group symmetry are formally broken in the construction of $\sutwo$ invariant PEPS which we described in Sec.~\ref{sec:SpinSingletPEPS}. In the present section we examine under which conditions this formal breaking of translation invariance or point group symmetry manifests at the physical level and has consequences for the transformation properties of the state. Since this question arises also in one-dimensional tensor networks we discuss the simpler MPS in parallel to two-dimensional PEPS throughout this section.

We begin with a statement of the problem in subsection~\ref{sec:SymmetryCompatibilityOverview} and a summary of our findings in subsection~\ref{sec:SymmetryCompatibilityResults} before moving to a detailed analysis in subsection~\ref{sec:SymmetryCompatibilityProof}. The section concludes with the discussion of two simple MPS examples in subsection~\ref{sec:SymmetryCompatibilityExamples}.

\subsection{Statement of problem and assumptions\label{sec:SymmetryCompatibilityOverview}}

As described in Sec.~\ref{sec:SpinSingletPEPS}, $\sutwo$ invariant PEPS (also referred to as spin-singlet PEPS) can be defined via local projection maps which intertwine between the physical spin representation and the tensor product of representations for the virtual spins (see Eq.~\eqref{intertwiner}). If the tensor network is contracted with identities $\omega_{ij} = \delta _{ij}$ on the bonds, $\sutwo$ invariance requires that nearest-neighbour virtual spins transform in opposite representations $(\rho_v,\rho^*_v)$ under $\sutwo$. For an explicitly translation-invariant PEPS with $\omega_{ij} = \delta _{ij}$ this implies that the unique local projection map cannot transform straightforwardly under reflections and rotations since its virtual spins possess non-identical $\sutwo$ representations.

However, the self-conjugacy of $\sutwo$ representations permits the rewriting of a virtual nearest-neighbour pair $(\rho_v,\rho^*_v)$ connected by $\omega_{ij} = \delta _{ij}$ as a pair of spins transforming in identical representations $(\rho_v,\rho_v)$ and connected by a virtual singlet $\omega_{ij} = Y_{ij}$ (see Sec.~\ref{sec:SpinSingletPEPS}). One can therefore consider spin-singlet PEPS with the same representation $\rho_v$ for every virtual spin, a single local projection map
\begin{equation}\label{TensorMapWithSinglets}
\amap: \rho_v^{\otimes 4} \rightarrow \rho_{phys}
\end{equation}
valid for every lattice site and singlets $\omega_{ij} = Y_{ij}$ on the bonds. Here, the local projection map is defined on four identical virtual spins and can therefore transform in a simple way under $C_{4v}$, for instance in a one-dimensional representation (Eq.~\eqref{PointGroupLocalTensor}). The space group transformations of the PEPS are then determined both by the point group representation of $\amap$ and by the space group transformations of the orientation pattern for the virtual singlets, where both contributions can be analyzed separately. This is the approach we follow here. Let us now specify our assumptions and notations.

For the rest of this section, $\amap$ denotes a local tensor map of the form~\eqref{TensorMapWithSinglets} with basis entries denoted as $\ahat^s_{lurd}$. As specified in Eq.~\eqref{intertwiner} $\amap$ acts as an intertwiner between the four virtual spins transforming in identical $\sutwo$ representations $\rho_v$ and the physical spin with representation $\rho_{phys}$. We assume that the physical particle has a well-defined spin $\rho_{phys}=s_{phys}$ whereas the virtual representation
\begin{equation}
\rho_v=\bigoplus_{\alpha=1}^ns_{\alpha}
\end{equation}
is a direct sum of $n$ irreducible representations of spin $s_\alpha$ for $\alpha=1,\dotsc,n$. As in Sec.~\ref{sec:PEPSBasics} we denote by 
\begin{equation}
Y=\bigoplus_{\alpha=1}^ne^{i\pi \rho_{s_{\alpha}}(S^y)}
\end{equation}
the virtual spin flip operator with square 
\begin{equation}
Z= Y^2 = \bigoplus_{\alpha=1}^ne^{2\pi i\rho_{s_{\alpha}}(S^z)}
\end{equation}
that is equal to $+1$ ($-1$) on integer (half-integer) representations. Moreover we demand that the local tensor map changes in a simple manner under point group operations, \ie under permutations of its virtual spins. Specifically, we assume that $\amap$ transforms either in a one-dimensional representation $\sigma$ as defined in Eq.~\eqref{PointGroupLocalTensor} or else that its real and imaginary part transform in one-dimensional representations $\sigma_1$ and $\sigma_2$, respectively (\cf Eq.~\eqref{A1+iA2LocalTensor} and the subsequent discussion). In the latter case, we abbreviate the transformation of $\amap$ as $\sigma = \sigma_1 + i \sigma_2$. For MPS the point group $C_2$ has trivial and fundamental representations denoted by boldface letters $\mathbf{A}$ and $\mathbf{B}$ which are even and odd under inversion, respectively.  

\begin{center}
	\begin{figure}[htb]
		\begin{center}
			\includegraphics[width= 0.99 \linewidth]{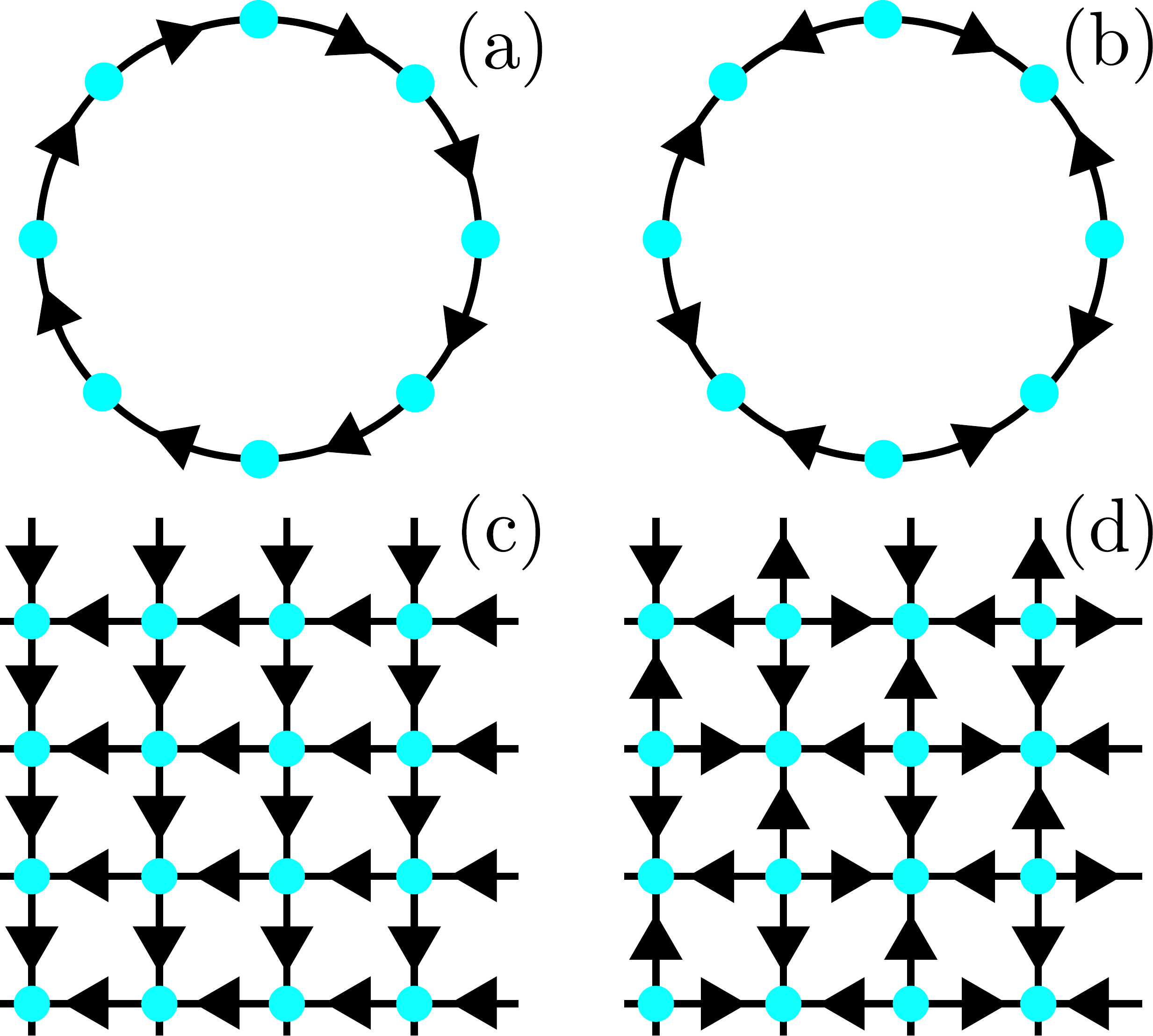}
		\end{center}
		\caption{Two different singlet orientation patterns for MPS in (a), (b) and PEPS in (c) and (d). The patterns in (a) and (c) are invariant under translation by one lattice site but not under point group operations, whereas the patterns (b) and (d) are invariant under the lattice point group but have an enlarged unit cell with respect to translation.}
		\label{fig::psi1psi2} 
	\end{figure}
\end{center}

An $\sutwo$ invariant PEPS is obtained from the local projection map $\amap$ by placing virtual singlets $w_{ij}=Y_{ij}$ on nearest-neighbour bonds. When $\ket{\omega}$ possesses an orientation (see Sec.~\ref{sec:SpinSingletPEPS}), the PEPS is well-defined only once we specify an orientation pattern for the singlets on all bonds. It is not possible to choose an orientation pattern for the square lattice that is simultaneously invariant under translations by one lattice site and invariant under rotations and reflections about lattice sites. In this construction either translation invariance or point group symmetry are therefore formally broken at the virtual level. We consider two natural choices for orientation patterns that are either translation invariant (see Fig.~\ref{fig::psi1psi2}(a) and (c)) or point group invariant (see Fig.~\ref{fig::psi1psi2}(b) and (d)). The latter pattern has an enlarged unit cell of 2 ($2\times 2$) sites in one dimension (two dimensions) and therefore applies only to lattices with an even number of sites in every direction and sub-lattices $\Lambda_{A(B)}=\{\sum_{i}n_i\mathbf{e}_i|\sum_i n_i \text{ even (odd)}\}$, where $i$ runs from $1$ to the number of spatial dimensions. We denote the PEPS and MPS derived from these two orientation patterns by $\psione$ and $\psitwo$, respectively. 

\begin{center}
	\begin{figure}[htb]
		\begin{center}
			\includegraphics[width= 0.99 \linewidth]{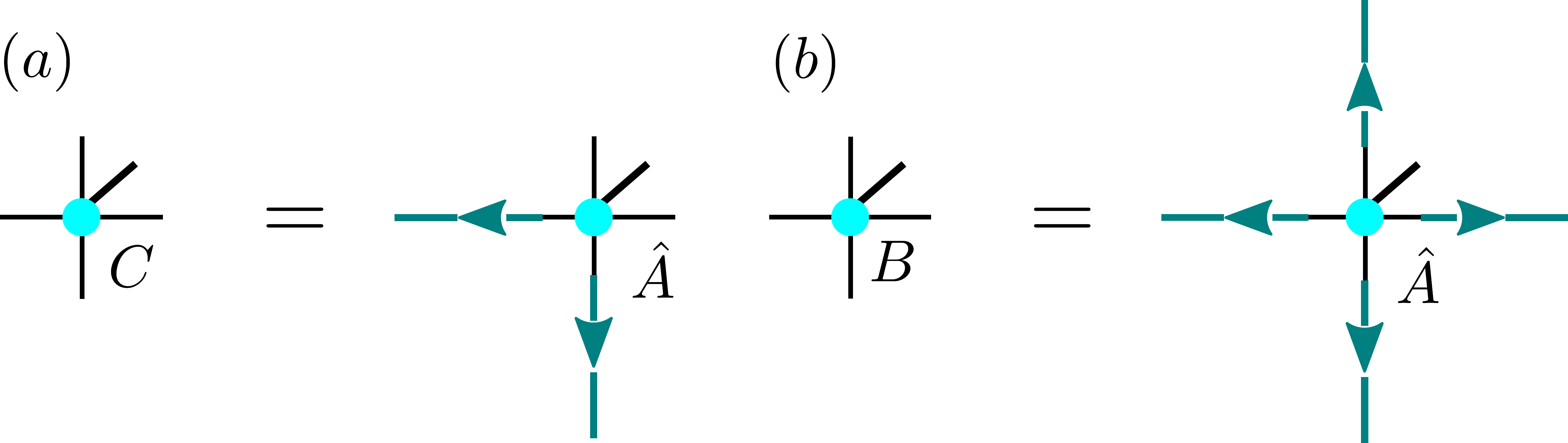}
		\end{center}
		\caption{Two different ways of absorbing oriented singlets into PEPS local tensors leading to (a) an explicitly translation invariant state or (b) a manifestly point group symmetric PEPS.}
		\label{fig::A_contracted} 
	\end{figure}
\end{center}

For future reference we want to express the states $\psione$ and $\psitwo$ in terms of tensor networks with identities $\delta_{ij}$ on all virtual bonds as in Eq.~\eqref{PEPSBasisCoefficient}. The state $\psione$ is computed from an explicitly translation invariant network with  a single local tensor
\begin{subequations}
	\label{DefC}
	\begin{gather}
	\text{MPS}:\quad   C^s_{lr}= \sum_{l'} Y_{l'l}\ahat^s_{l'r},\\
	\text{PEPS}:\quad  C^s_{lurd}= \sum_{l'd'} Y_{l'l}Y_{d'd} \ahat^s_{l'urd'}.
	\end{gather}
\end{subequations}
that is obtained by absorbing the virtual singlet on every left (left and down) virtual leg for MPS (PEPS), respectively (see Fig.~\ref{fig::A_contracted}(a) for PEPS). The local tensor $C$ does not generally transform under point group operations in the representation $\sigma$ as defined above. On the other hand, $\psitwo$ has a sublattice structure with local tensors $A$ ($B$) for sites on $\Lambda_A$ ($\Lambda_B$), where $A^s_{lurd} = \ahat^s_{lurd}$ and $B$ is obtained by absorbing the virtual singlets on all virtual legs  (see Fig.~\ref{fig::A_contracted}(b) for PEPS) such that
\begin{subequations}
	\begin{gather}
	\text{MPS}: \quad B^s_{lr}=\sum_{l'r'}Y_{l'l}\ahat^s_{l'r'}Y_{r'r},\\
	\text{PEPS}: \quad B^{s}_{lurd}=\sum_{l'u'r'd'}Y_{l'l}Y_{u'u}Y_{r'r}Y_{d'd}\ahat^s_{l'u'r'd'}.
	\end{gather}
\end{subequations}   
Due to the $\sutwo$ symmetry of the local projection map the local tensor $B$ can also be obtained by applying a physical spin flip $e^{i\pi \rho_{phys}(S^y)}$ to the tensor $A$. Therefore both $A$ and $B$ transform under $C_{4v}$ in the representation $\sigma$.

The space group transformations of the states $\psione$ and $\psitwo$ are determined both by the point group representation $\sigma$ of the local tensor map and the transformation of the respective singlet orientation pattern. We denote by $\Sigma (\sigma)$ the contribution of $\amap$ to the PEPS and MPS point group transformations. $\Sigma (\sigma)$ is a one-dimensional representation or consists of independent one-dimensional representations for the real and imaginary parts and generally also depends on the system size (\cf Sec.~\ref{sec:SpaceGroup}). The singlet orientation pattern for the state $\psitwo$ is invariant under rotations and reflections about lattice sites such that $\psitwo$ transforms under point group operations as $\Sigma (\sigma)$. However, $\psitwo$ is not manifestly translation invariant. On the other hand, the orientation pattern for $\psione$ is invariant under translations but not point group transformations. The state $\psione$ is therefore manifestly translation invariant but generally transforms under point group operations in a manner different from $\Sigma (\sigma)$.

The questions we want to answer are therefore threefold:

\begin{enumerate}
	\item How does the translation invariant state $\psione$ transform under point group operations?
	
	\item How does the point group symmetric state $\psitwo$ transform under lattice translations?
	
	\item What is the relation between the two states $\psione$ and $\psitwo$ when both are well-defined, \ie when the number of lattice sites in every direction is even?
\end{enumerate}

\subsection{Summary of results\label{sec:SymmetryCompatibilityResults}}

The answers to the questions above will depend on the system size and also on the physical and virtual spins $s_{phys}, s_{\alpha}$. $N$ denotes the total system size, \ie the length of the chain for MPS and $N=N_hN_v$ for PEPS. Let us recall that $N$ should be even for half-integer spins and $\Psi_2$ is defined only if both $N_h$ and $N_v$ are even. We will distinguish three cases of interest to us: 
\begin{enumerate}[label =Case~\Roman*, align=parleft,leftmargin=*  ]
	\item The virtual spins are purely integer such that $s_{phys}$ is integer by the $\sutwo$ fusion rules.
	
	\item The virtual spins are purely half-integer such that $s_{phys}$ is integer.
	
	\item The virtual spins are mixed integer and half-integer and $s_{phys}$ is half-integer such that the local tensor possesses the virtual $\mathbb{Z}_2$ symmetry defined in Eq.~\eqref{VirtualZ2Symmetry}.
\end{enumerate}
A fourth case exists where the physical spin is integer and the virtual spins are mixed integer and half-integer. However, this situation is less relevant than Case III since the latter constitutes the only possibility to build PEPS for half-integer physical spin whereas for integer spin Case I and II provide more natural options. As the results for both cases are identical up to sign factors we focus here on Case I-III. We found the following:
\begin{enumerate}
	\item Point group transformations of $\psione$:
	\begin{enumerate}[label =Case~\Roman*, align=parleft,leftmargin=*  ]
		\item $\psione$ transforms as $\Sigma (\sigma)$.
		
		\item $\psione$ transforms as $\Sigma (\sigma)$ for $N$ even and as $\mathbf{B} \otimes \Sigma(\sigma)$ ($\mathbf{B}_2 \otimes \Sigma(\sigma)$) for MPS (PEPS) with $N$ odd.
		
		\item $\psione$ transforms as $\Sigma (\sigma)$ on chains of length $N\in 4\mathbb{N}$, as $\mathbf{B} \otimes \Sigma(\sigma)$ on chains of length $N\in 4\mathbb{N}+2$ and it vanishes on odd-length chains. In two dimensions it transforms as $\Sigma (\sigma)$ on even-by-even tori and vanishes if both $N_h,N_v$ are odd. It transforms in a two-dimensional representation on non-quadratic tori with $N_h$ even and $N_v$ odd such that it is mapped to the state $\psione_h$ ($(-1)^{N_h/2}\psione$) with a horizontal flux line under the horizontal (vertical) mirror, and analogously for $N_h$ odd and $N_v$ even.
	\end{enumerate}
	
	\item Translation of $\psitwo$:
	
	\begin{enumerate}[label =Case~\Roman*, align=parleft,leftmargin=*  ]
		\item $\psitwo$ is translation invariant.
		
		\item $\psitwo$ is translation invariant.
		
		\item Translation changes $\psitwo$ by a phase $(-1)^{N/2}$ that is non-trivial on chains of length $N\in4\mathbb{N}+2$ but which is always trivial in two dimensions since $N\in4\mathbb{N}$ for even-by-even tori.
	\end{enumerate}
	
	\item Relation of $\psione$ and $\psitwo$:
	
	\begin{enumerate}[label =Case~\Roman*, align=parleft,leftmargin=*  ]
		\item $\psione = \psitwo$.
		
		\item $\psione = (-1)^{N/2} \psitwo$ for MPS and $\psione = \psitwo$ for PEPS.
		
		\item The two states possess different $\mathbb{Z}_2$ fluxes around non-contractible loops if the system size in at least one direction is not divisible by four. For MPS
		\begin{subequations}
			\label{Eq:EquivalenceMPS}
			\begin{align}
			N\in 4\mathbb{N}:\quad& \ket{\psi_1}=\ket{\psi_2},\\
			N\in 4\mathbb{N}+2:\quad& \ket{\psi_1}=\ket{\psi_2}_Z
			\end{align}    
		\end{subequations}
		and for PEPS
		\begin{subequations}\label{EquivalenceTorus}
			\begin{align}
			N_h,N_v\in 4\mathbb{N}:&\quad \psitwo = \psione,\\
			N_{h(v)}\in 4\mathbb{N},\, N_{v(h)}\in 4\mathbb{N}+2:&\quad \psitwo = \psione_{h(v)},\\
			N_h,N_v\in 4\mathbb{N}+2:&\quad \psitwo = -\psione_{h,v}.
			\end{align}
		\end{subequations}
		where the states with flux insertions are as defined in Sec.~\ref{sec:VirtualSymmetry}.
	\end{enumerate} 
\end{enumerate}

\subsection{Proofs\label{sec:SymmetryCompatibilityProof}}

\subsubsection{Point group transformations of $\psione$}

The deviations of the point group transformations of $\psione$ from $\Sigma (\sigma)$ are caused by the non-trivial transformation of the corresponding singlet orientation pattern displayed in Fig.~\ref{fig::psi1psi2}(a) and (c). The flipping of an arrow in this singlet orientation pattern corresponds to the insertion of a matrix $(Y^T)^{-1}Y = Z$ on that virtual bond. 

Case I: $Z=\mathbbm{1}$ such that the flipping of arrows does not manifest at the physical level. 

Case II: $Z=-\mathbbm{1}$ such that the flipping of arrows results in an overall phase $(-1)^N$ for horizontal and vertical reflections as well as rotations. Diagonal reflections leave the singlet pattern invariant or cause a trivial phase $(-1)^{2N}$.

Case III: $Z\neq \pm \mathbbm{1}$ such that rotations and reflections cause the insertion of $\mathbb{Z}_2$ fluxes. On one-dimensional chains reflection causes the insertion of $N$ matrices $Z$ that cancel pairwise due to the virtual $\mathbb{Z}_2$ symmetry given by Eq.~\ref{VirtualZ2Symmetry}. The $(-1)$ factor in this equation leads to an overall sign of $(-1)^{N/2}$ for the MPS (note that $N$ is even for half-integer $s_{phys}$). For PEPS, the horizontal mirror and rotation by $\pi/2$ (vertical mirror and rotation by $-\pi/2$) cause the insertion of a $Z$-matrix on every vertical (horizontal) bond and therefore of $N_v$ horizontal ($N_h$ vertical) flux lines. These can be rearranged to obtain the transformation properties stated above. Note that $\psione$ vanishes if the total number of sites is odd since in this case the half-integer physical spins cannot fuse to an $\sutwo$ invariant state.  

\subsubsection{Translation of $\psitwo$}

Translation of $\psitwo$ by one lattice site exchanges the two sublattices and acts on the state as a staggered physical spin flip
\begin{equation}\label{SublatticeExchange}
\bigotimes_{\mathbf{x}\in\Lambda_A} e^{i\pi \rho_{phys}(S^y)} \bigotimes_{\mathbf{x}\in\Lambda_B} e^{-i\pi \rho_{phys}(S^y)}.
\end{equation}
The physical spin flip is related to its inverse as $e^{i\pi \rho_{phys}(S^y)} = \pm e^{-i\pi \rho_{phys}(S^y)}$ for integer (half-integer) $s_{phys}$ and $\psitwo$ is invariant under a global spin flip. It is therefore invariant under lattice translations in Case I and Case II but changes by a phase $(-1)^{N/2}$ in Case III. This phase is always trivial for PEPS but can be relevant in MPS such as for the Majumdar-Ghosh chain, see Sec.~\ref{sec:SymmetryCompatibilityExamples}.

\subsubsection{Relation between $\psione$ and $\psitwo$}

\begin{center}
	\begin{figure}[htb]
		\begin{center}
			\includegraphics[width=\linewidth]{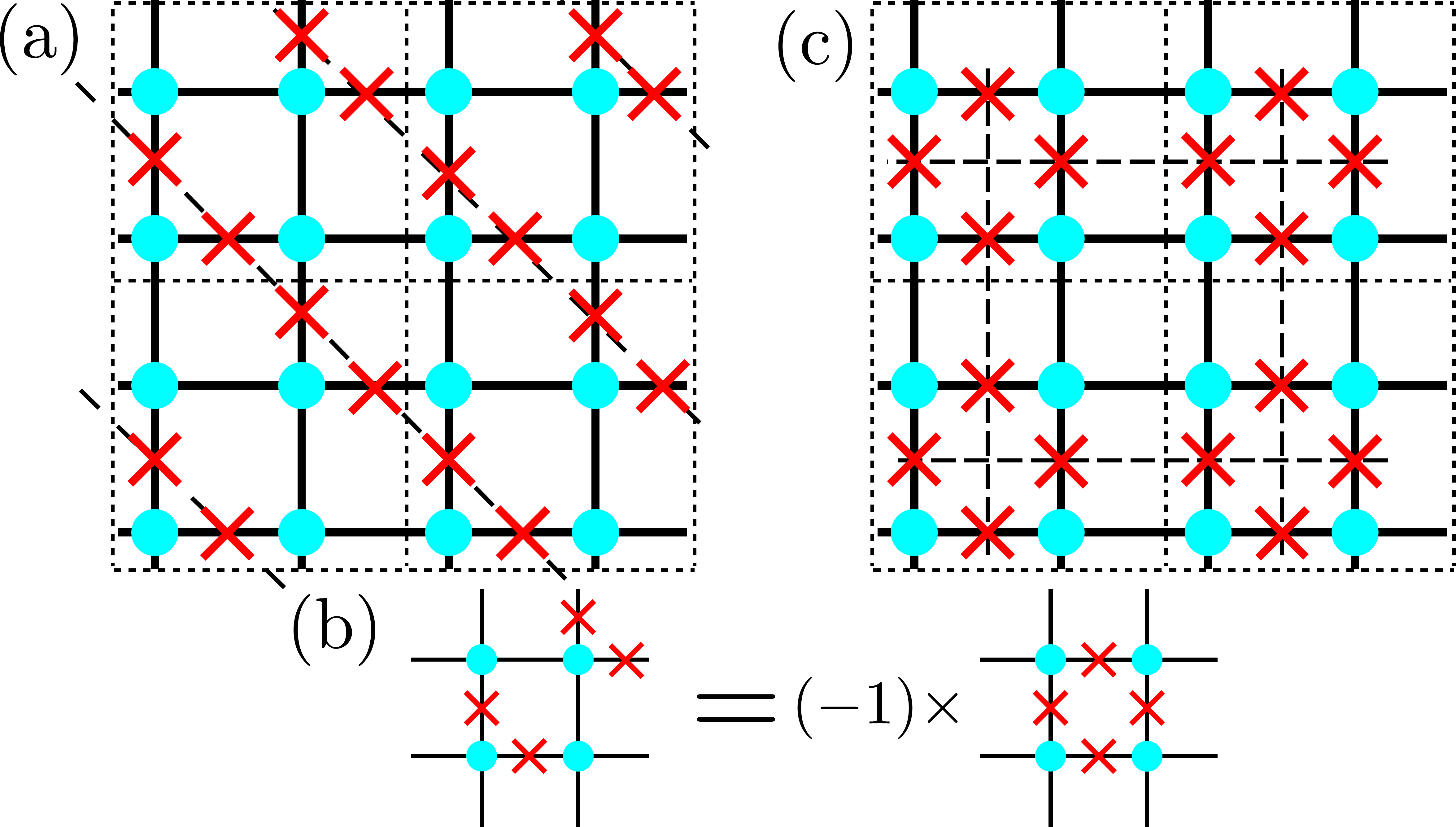}
		\end{center}
		\caption{(a) Schematic representation of the difference between the singlet orientation patterns defining $\psione$ and $\psitwo$. Whenever the direction of a singlet has to be exchanged this corresponds to the insertion of a matrix $Z$ on the corresponding bond. The inserted $Z$-matrices are represented by red crosses while the physical legs are suppressed in the interest of readability. The dotted lines delimit patches on which the graphical equation given in (b) will be applied whereas the dashed lines indicate flux lines. (b) Using the $\mathbb{Z}_2$ virtual symmetry, the two outer crosses can be moved to the interior links. (c) As a result of applying (b) to the 4 squares on (a), a PEPS with a flux line along every other row and every other column is obtained.}
		\label{fig::inequivalenceps1ps2} 
	\end{figure}
\end{center}

The singlet orientation patterns defining $\psione$ and $\psitwo$ are related by the flipping of the arrow direction on every other nearest-neighbour bond corresponding to the insertion of the matrix $(Y^T)^{-1}Y=Z$ on that link (see Fig.~\ref{fig::inequivalenceps1ps2}(a)). Case I: $Z=\mathbbm{1}$ such that the two states are identical. Case II: $Z=-\mathbbm{1}$ such that the states are related by a phase $(-1)^{N/2}$ for MPS and $(-1)^{N}$ for PEPS. 

For Case III the $Z$-insertions can be rearranged using the virtual $\mathbb{Z}_2$ symmetry of every local tensor, see Fig.~\ref{fig::inequivalenceps1ps2}(b) for PEPS. In one dimension, the total number of such $Z$-insertions is given by $N/2$ and thus even (odd) on chains of length $N\in 4\mathbb{N}$ ($N\in 4\mathbb{N}+2$), giving the relation~\eqref{Eq:EquivalenceMPS}. In two dimensions, the $Z$-insertions on every second horizontal and vertical link can be rearranged to obtain a network with a $\mathbb{Z}_2$ flux line wrapping around every other horizontal and vertical line through the centers of plaquettes such that there are a total of $N_v/2$ ($N_h/2$) horizontal (vertical) flux lines, respectively (see Fig.~\ref{fig::inequivalenceps1ps2}(c)). Since flux lines around the same non-contractible loop cancel pairwise one obtains the result~\eqref{EquivalenceTorus}.

\subsection{Examples\label{sec:SymmetryCompatibilityExamples}}

In this subsection we illustrate the previous discussion with two examples of spin-chain ground states which have an exact MPS representation: the AKLT state~\cite{affleck1987rigorous} and the Majumdar-Ghosh state~\cite{majumdar1969next}.

\subsubsection{AKLT-type MPS}

\begin{center}
	\begin{figure}[htb]
		\begin{center}
			 \includegraphics[width=\linewidth]{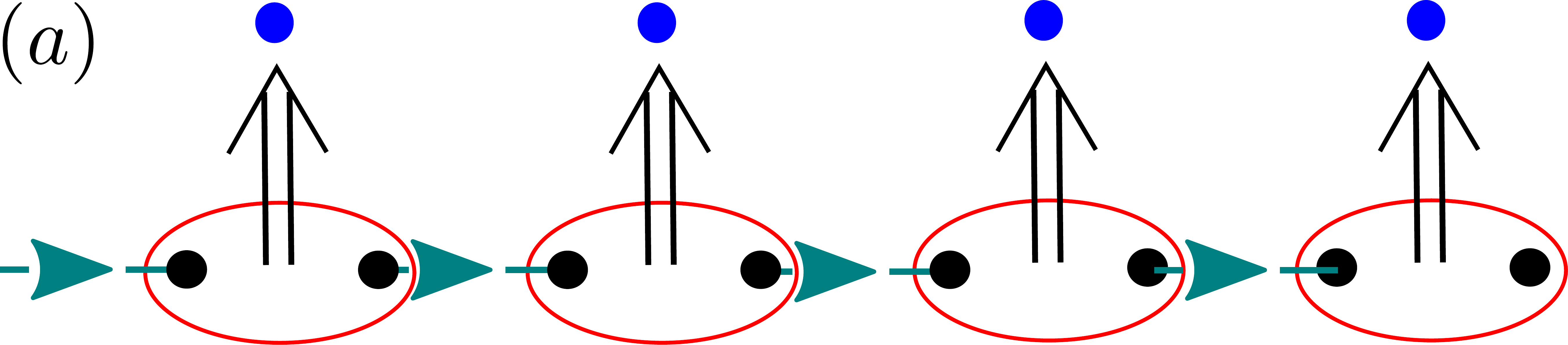}\\ \vspace*{0.7 cm}
			 \includegraphics[width=\linewidth]{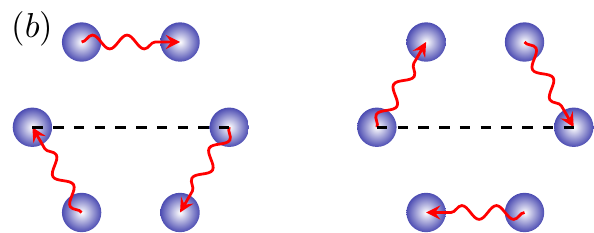}
		\end{center}
		\caption{(a): Schematic representation of the AKLT state: Two spin-$s$ virtual spins are projected onto the maximal spin sector $2s$ at each site and are connected to their nearest neighbours by a singlet state. Depending on $s$, the singlet state is oriented (case shown here) or not. (b): Schematic description of the two degenerate Majumdar-Ghosh states $\ket{\chi_1}$ and $\ket{\chi_2}$ where the physical singlets are represented by red arrows. While they are directly related by translation, a reflection around the dashed line exchanges the two states but also flips the anti-symmetric singlets giving a sign $(-1)^{N/2}$.}
		\label{fig::examples_MPS} 
	\end{figure}
\end{center}

A general AKLT MPS is constructed by choosing an irreducible representation $\rho_v = s$ for all virtual legs and by projecting the two virtual spins corresponding to a physical site onto the maximal spin $s_{phys}=2s$. Thus integer (half-integer) $s$ corresponds to Case I (Case II) of subsection~\ref{sec:SymmetryCompatibilityResults}, respectively. This construction is sketched in Fig.~\ref{fig::examples_MPS}(a). The resulting local tensor map is inversion symmetric, \ie $\ahat^{s}_{lr}= \ahat^{s}_{rl}$. On the other hand, the singlets created by the operator $Y=\rho_v(e^{i\pi S_y})$ are symmetric (anti-symmetric) for integer (half-integer) virtual spin $s$. The local tensor of the translation invariant state $\psione$ is therefore symmetric (anti-symmetric) up to a local basis change
\begin{equation}\label{TrafoAKLT}
(C^s)^T =  (-1)^{2s}\, Y^{-1} C^s Y.
\end{equation}
The MPS $\psione$ therefore transforms under the point group in the representation $\mathbf{A}$ unless $s$ is half-integer and $N$ is odd in which case it transforms as $\mathbf{B}$. Interestingly, the sign in Eq.~\eqref{TrafoAKLT} is a manifestation of the topologically non-trivial (trivial) nature of the AKLT state for odd (even) spin $s_{phys}$ as pointed out in Ref.~\cite{2012PhRvB..85g5125P}. Since the singlet orientations contribute only an overall sign that is trivial for even $N$, $\psitwo$ is always both translation-invariant and equal to $\psione$.

\subsubsection{Majumdar-Ghosh MPS}

The Majumdar-Ghosh state is a spin-$\frac{1}{2}$ valence bond state where every physical spin forms a singlet with one of its nearest neighbors (hence the number of sites must be even). On a periodic chain there are two possible states $\ket{\chi_1},\ket{\chi_2}$ sketched in Fig.~\ref{fig::examples_MPS}(b) for which the first spin forms a singlet either with the second one (on the right) or with the last one (on the left). Both translation $T$ and inversion $\mathcal{M}$ around sites exchange the two states,  $\ket{\chi_2} = T\ket{\chi_1}$ and $\mathcal{M}\ket{\chi_1}= (-1)^{N/2}\ket{\chi_2}$. Here, the sign counts the anti-symmetric physical singlets flipped by inversion.

Since translation symmetry is broken in the valence bond description, a translation invariant MPS exists only for linear superpositions of $\ket{\chi_1}$ and $\ket{\chi_2}$. This MPS has virtual $\sutwo$ representations $\rho_v=0\oplus \frac{1}{2}$ and the local tensor map $\ahat^{s}_{lr}=\delta_{0,l}\delta_{s,r}+\delta_{0,r}\delta_{s,l}$. Thus this example belongs to Case III of subsection~\ref{sec:SymmetryCompatibilityResults}. It is fully specified only once we choose an orientation pattern for the virtual singlets. Indeed, the orientation of a physical singlet between two sites in the resulting state corresponds to the orientation of the virtual singlet of the corresponding bond. The translation-invariant and inversion-symmetric MPS are therefore given by the linear superpositions $\psione = \ket{\chi_1} + T\ket{\chi_1}$ and $\psitwo = \ket{\chi_1} +\mathcal{M}\ket{\chi_1}$, respectively. These two states are the same for $N\in4\mathbb{N}$ whereas $T\ket{\chi_1}= - \mathcal{M}\ket{\chi_1}$ for $N\in4\mathbb{N} + 2$ due to the odd number of physical singlets. In the tensor network language, we can account for this by multiplying one bond with the matrix $Z$ which adds an overall phase $-1$ when this bond is crossed by a physical singlet. Therefore $\psitwo = \psione_Z$ for $N\in4\mathbb{N} + 2$ as shown in Eq.~\ref{Eq:EquivalenceMPS}.

\section{Symmetries of chiral spin liquid PEPS\label{sec:0Plus12PEPS}}

From now on until the end of this paper we study the chiral PEPS for spin-$\frac{1}{2}$ on the square lattice that was introduced in Ref.~\cite{Poilblanc2015PRB_chiralPEPS} and subsequently studied in Ref.~\cite{Poilblanc2016PRB_chiralPEPS}. Unlike the chiral PEPS from Refs.~\cite{wahl2013projected, Wahl2014PRB, Yang2015chiral} this state is not defined in terms of free fermions or Gutzwiller projections thereof. Instead, the PEPS is constructed from interacting bosonic spins by considering the most general local tensor map satisfying certain symmetry conditions: $\mathbf{A}_1+i\mathbf{A}_2$ symmetry under $C_{4v}$ and invariance under simultaneous physical and virtual $\sutwo$ rotations, where the virtual spins are assumed to transform in the representation $\rho_v=\mathbf{0}\oplus \mathbf{\frac{1}{2}}$. One thereby obtains a family of spin-singlet states parametrized by three real numbers $\lambda_1,\lambda_2,\lambda_c$. These spin liquids are deformations of the PEPS for the nearest-neighbour RVB state~\cite{Schuch2012PRB_RVB,Poilblanc2012PRB_RVB}, where additional terms with amplitudes $\lambda_2$ and $\lambda_c$ in the local tensor generate long-range singlets~\cite{PhysRevLett.111.037202}. They provide good variational states for the square lattice Heisenberg model with additional chiral cyclic plaquette terms~\cite{PhysRevB.96.121118}. Moreover, for $\lambda_c=0$ it was shown that this PEPS can be either critical or in the $\mathbbm{Z}_2$ spin liquid phase \cite{PhysRevB.97.161107}.

The ES for the chiral PEPS presented in Refs.~\cite{Poilblanc2015PRB_chiralPEPS,Poilblanc2016PRB_chiralPEPS} resembles the spectrum of the chiral conformal field theory (CFT) $\mathfrak{su}(2)_1$. However, this correspondence is not perfect. Firstly, the ES for the state with $\mathbb{Z}_2$-even boundary conditions exhibited unambiguous chiral features only after the momentum was projected onto the region between $0$ and $\pi$. Secondly, the ES for the state with $\mathbb{Z}_2$-odd boundary conditions displayed two identical modes shifted in momentum by $\pi$. Interpreted as the two sectors of the CFT $\mathfrak{su}(2)_1$, these spectra therefore do not give the expected conformal weight of $h = 1/4$~\cite{francesco2012conformal}. Moreover, the ES of the state with a flux line along the cylinder was found not to display any chiral features.  

In this section we are going to conduct a comprehensive analysis of the symmetries possessed by the chiral spin liquid PEPS which follow from the special form of its local projection map. This understanding will permit us to explain some of the discrepancies between its ES and the CFT $\mathfrak{su}(2)_1$. In particular we show that the $\mathbb{Z}_2$-even ES is chiral in the entire Brillouin zone for the explicitly translation invariant PEPS defined in the previous section. Moreover we prove that the two branches in the $\mathbb{Z}_2$-odd ES follow from a dressed mirror symmetry. In Sec.~\ref{sec:ZeroLambdaOne}, this identification will allow us to resolve this issue by considering states which break this symmetry and have only a single branch in the ES. We also compute the momentum-resolved ES of the PEPS with a horizontal flux line and show that it has some chiral features even though they do not appear linked to a simple CFT.

After formally defining the chiral PEPS in Sec.~\ref{sec:PEPSDef} we analyze the symmetries of its transfer matrix and the generic form of its spectrum in Sec.~\ref{sec:SymmetriesTransferMatrix} and Sec.~\ref{sec:SpectrumTransferMatrix}, respectively. We provide the same analysis for the transfer matrix with a horizontal flux line in Sec.~\ref{sec:TransferMatrix_string}. Finally, we investigate the ES of the corresponding fixed points in Sec.~\ref{sec:SymmetriesES}.

\subsection{Definition\label{sec:PEPSDef}}

We study a spin liquid PEPS for particles with spin $s_{phys}=\frac{1}{2}$ on a square lattice. The state has bond dimension $D=3$ and virtual $\sutwo$ representations 
\begin{equation}
\label{VirtualSpace}
\rho_v=\mathbf{0}\oplus \mathbf{\frac{1}{2}}.
\end{equation}
The local projection map is an intertwiner of $\sutwo$ representations and transforms in the representation $\mathbf{A}_1+i\mathbf{A}_2$ of $C_{4v}$. It therefore possesses the virtual $\mathbb{Z}_2$ symmetry from Eq.~\eqref{VirtualZ2Symmetry}. Specifically, one chooses
\begin{equation}
\label{defA}
\amap = \lambda_1\mathcal{P}(\mathbf{A}_1^{\varphi}) + \lambda_2 \mathcal{P}(\mathbf{A}_1^{3\varphi}) + i\lambda_c \mathcal{P}(\mathbf{A}_2^{3\varphi}),
\end{equation}
where $\mathcal{P}$ are projections on irreducible $C_{4v}$ and spin-$\frac{1}{2}$ representations in the tensor product $\rho_v^{\otimes 4}$. The coefficients $\lambda_1,\lambda_2$ are real for the $\mathbf{A}_1$-representations whereas $i\lambda_c$ is purely imaginary for the representation $\mathbf{A}_2$. The superscripts $\varphi,3\varphi$ refer to the transformation under a local $\uone$ action described below. The concrete form of the projection map is given in the appendix.

We can equivalently define the local projection map~\eqref{defA} as a superposition of the spin-$\frac{1}{2}$ $\mathbf{B}_{1,2}$ representations in $\rho_v^{\otimes 4}$~\cite{Poilblanc2015PRB_chiralPEPS}. Indeed, due to the virtual $\mathbb{Z}_2$ symmetry the two sets of local tensors corresponding to the $\mathbf{A}_{1,2}$ and $\mathbf{B}_{1,2}$ representations are related by a local gauge transformation and therefore define the same PEPS on tori and cylinders~\footnote{Given a local tensor transforming in the $\mathbf{A}_{1,2}$ representation, one obtains a tensor transforming as $\mathbf{B}_{1,2}$ by multiplying either both horizontal or both vertical virtual legs with the matrix $Z$ from Eq.~\eqref{VirtualZ2Symmetry}.}.

Since $\rho_v$ is a sum of spin representations it carries a $\uone$ action
\begin{equation}
\label{UOneAction}
U(\varphi): \ket{v}+\ket{w}\in \mathbf{0}\oplus \mathbf{\frac{1}{2}}\mapsto \ket{v}+e^{i\varphi}\ket{w}
\end{equation}
which modifies the relative phase of vectors in the spin-$0$ and spin-$\frac{1}{2}$ subspaces. States in the representations $\mathbf{A}_1^{\varphi},\mathbf{A}_1^{3\varphi}, \mathbf{A}_2^{3\varphi}$ that define the local tensor have different eigenvalues $\varphi$ ($3\varphi$) under this group action such that a simultaneous group action on all four virtual legs changes the parameter $\lambda_1$ relative to $\lambda_{2,c}$,
\begin{equation}
\label{UOneTensor}
\amap(\lambda_1,e^{2i\varphi}\lambda_2,e^{2i\varphi}\lambda_c)=e^{-i\varphi}\amap(\lambda_1,\lambda_2,\lambda_c) \circ U(\varphi)^{\otimes 4}.
\end{equation}

The transformation~\eqref{UOneTensor} implies that on an even-by-even torus the PEPS with parameters $(\lambda_1,\lambda_2,\lambda_c)$ is equal to the state with parameters $(\lambda_1,-\lambda_2,-\lambda_c)$. In particular it is real if either $\lambda_2$ or $\lambda_c$ vanishes. For a proof we multiply all four virtual legs corresponding to tensors on the sub-lattice $\Lambda_A$ ($\Lambda_B$) with $U(\pi/2)$ ($U(-\pi/2)$) such that the two transformations cancel each other on every bond and the tensor network remains invariant. On the other hand, according to Eq.~\eqref{UOneTensor} the $\uone$ transformations change the parameters of every local tensor as $(\lambda_1,\lambda_2,\lambda_c) \mapsto(\lambda_1,-\lambda_2,-\lambda_c)$ up to a phase which cancels on a patch of $2\times 1$ lattice sites.

\subsection{Transfer matrix symmetries\label{sec:SymmetriesTransferMatrix}}

In this section we study the symmetries of the transfer matrix $\Gamma$ of the translation invariant PEPS $\psione$ on cylinders of even width $N_v$. Our results are summarized in Table~\ref{tab:SymmetriesTransferMatrix}. Since the two transfer matrices are unitarily equivalent after blocking four columns up to the insertion of a horizontal flux line for $N_v\in4\mathbb{N}+2$ (see Sec.~\ref{sec:SymmCompatibility}), our results apply also to the point-group symmetric state $\psitwo$ after corresponding transformations of the quantum numbers. The notation $\mathcal{O}_1\otimes\mathcal{O}_2$ refers to a linear operator for a double-layer column of virtual spins that is a tensor product of two operators $\mathcal{O}_{1,2}\in\End(\hsl)$ acting on the ket-layer and bra-layer separately. We denote by $\mathbf{l}\equiv(l_0,\dotsc,l_{N_v-1})\in\{0,\dotsc,D-1\}^{N_v}$ a multi-index for the single-layer virtual space $\hsl$ and by $\mathbf{L} = (\mathbf{l},\mathbf{\tilde{l}})$ a multi-index for the double-layer space $\hsl\otimes\hsl$.

\begin{table}[t]
	\centering
	\begin{ruledtabular}
		\begin{tabular}{l|l|l|l}
			Operator & Symbol & Properties & \makecell{Quantum\\numbers} \\
			\hline\hline
			\makecell{Transfer\\matrix} & $\Gamma$ & \makecell{(Anti-)\\ Hermitian\\
			for $s^z_{dl}$\\(half-)integer} & $E$\\
			\hline
			$\sutwo$ spin& \makecell{$\sum_a\rhodl(S^a)^2$,\\$\rhodl(S^z)$} &Hermitian& $s_{dl},s^z_{dl}$ \\
			\hline
			$\mathbb{Z}_2$ charges& \makecell{$Z^{\otimes N_v}\otimes \mathbbm{1}^{\otimes N_v}$, \\ $\mathbbm{1}^{\otimes N_v} \otimes Z^{\otimes N_v}$} & \makecell{ Hermitian \& \\ unitary}& $Z_k, Z_b$ \\ 
			\hline
			\makecell{Translation} & $T^{dl}$ & unitary & $K^{dl}$ \\
			\hline\hline
			\makecell{Layer \\ inversion} & $I$ & anti-unitary & \makecell{$E\mapsto (-1)^{\epsilon} E$\\$K^{dl}\mapsto -K^{dl}$\\$Z_k\leftrightarrow Z_b$} \\
			\hline
			\makecell{Dressed\\mirror} & $\rdl$ & anti-unitary & \makecell{$E\mapsto (-1)^{\epsilon} E$\\$K^{dl}\mapsto K^{dl}+\epsilon \pi$}\\
		\end{tabular}
	\end{ruledtabular}
	\caption{Operators commuting with the transfer matrix $\Gamma$ which define quantum numbers (upper part) or which create multiplets (lower part) according to Eq.~\eqref{LayerInversionZ2} and Eq.~\eqref{MirrorTranslation} (we exclude the $\sutwo$ ladder operators). Here, $\epsilon = 0$ ($\epsilon = 1$) for the integer (half-integer) spin sector. The table is valid also for the transfer matrix $\Gammah$ with a horizontal flux line in both layers after substitution of the dressed translation operator $\tddl$ for $T^{dl}$.\label{tab:SymmetriesTransferMatrix}}
\end{table} 

\subsubsection{$\sutwo$ symmetry}

\begin{center}
 \begin{figure}[htb]
\begin{center}
\includegraphics[width=0.9\linewidth]{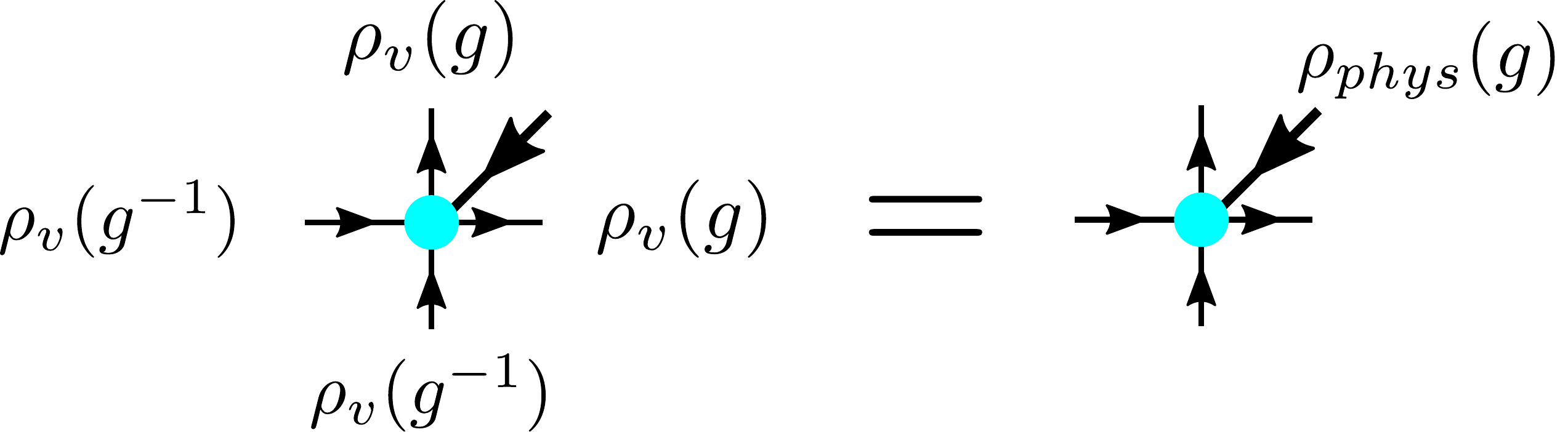}
\end{center}
\caption{Factorization of a physical spin rotation as a product of virtual $SU(2)$ transformations for the local tensor $C$ of the translation invariant PEPS $\psione$. The arrow directionality is a consequence of the singlet absorption pattern.}
\label{fig::PEPS_su2} 
\end{figure}
\end{center}

Spin rotations applied to the physical leg of the local tensor $C$ can be pushed to the virtual layer and factorise as a product of representations acting on every virtual leg individually. Due to the virtual spin flips contained in the definition~\eqref{DefC} of $C$, its left and down (up and right) virtual legs transform in the representation $\rho_v^*$ ($\rho_v$). This may be interpreted as an ingoing (outgoing) transformation $\rho_v(g^{-1})$ ($\rho_v(g)$) acting on the left and down (up and right) virtual legs of $C$ as sketched in Fig.~\ref{fig::PEPS_su2}. The complex conjugate tensor $C^*$ for the bra layer transforms under $\sutwo$ in a similar fashion but with conjugate representations $\rho_{phys}^*$ for the physical leg and $\rho_v$ ($\rho_v^*$) for the left and down (up and right) virtual legs. Hence, $\Gamma$ commutes with simultaneous $\sutwo$ transformations
\begin{equation}
\rhodl = \rhosl \otimes \rhosl ^*= \rho_v^{\otimes N_v}\otimes (\rho_v^*)^{\otimes N_v}
\end{equation}of all virtual spins in both layers.

\subsubsection{Virtual $\ztwo$ symmetry}\label{sec::z2sym}

The total charge $Z^{\otimes N_v}$ of the virtual $\mathbb{Z}_2$ symmetry is conserved independently in both layers of the transfer matrix (see Eq.~\eqref{VirtualZ2Symmetry}). We denote the eigenvalues of the ket-layer and bra-layer operators $Z^{\otimes N_v}\otimes \mathbbm{1}^{\otimes N_v}$ and $\mathbbm{1}^{\otimes N_v}\otimes Z^{\otimes N_v}$ by $Z_{k}$ and $Z_{b}$, respectively. One has
\begin{subequations}
\begin{gather}
Z^{\otimes N_v}=\rhosl(e^{2\pi i S^z})\\
Z^{\otimes N_v}\otimes Z^{\otimes N_v}=\rhodl(e^{2\pi i S^z})
\end{gather}
\end{subequations}
such that the configurations $(Z_{k},Z_{b}) \in\{(1,1), (-1,-1)\} $ ($\{(1,-1), (-1,1)\}$) correspond to integer (half-integer) double-layer virtual spin $\rhodl (S^z)$.

\subsubsection{Translation invariance}

The transfer matrix $\Gamma$ commutes with the unitary translation operator $T^{dl}\equiv  (T^{sl})^{\otimes 2}$ defined by 
\begin{equation}
T^{sl}_{\mathbf{l}\mathbf{r}}=\delta_{l_0,r_1}\delta_{l_1,r_2}\dotsm\delta_{l_{N_v-1},r_0}. 
\end{equation}
The single-layer (double-layer) translation operator satisfies $(T^{sl(dl)})^{N_v}=\mathbbm{1}$ such that the momenta are $K^{sl(dl)}=2\pi n/N_v$ with $n=0,\dotsc,N_v-1$.

\subsubsection{Layer inversion}

The transfer matrix is complex conjugated when the indices in its ket- and bra-layers are exchanged, \ie $\mathbf{L} = (\mathbf{l},\mathbf{\tilde{l}})\mapsto \mathbf{L'} = (\mathbf{\tilde{l}},\mathbf{l})$. Formally, $\Gamma$ commutes with an anti-unitary operator $I=\mathcal{C}\circ \tilde{I}$ where $\tilde{I}$ is a unitary double-layer operator with basis elements $\tilde{I}_{\mathbf{L}\mathbf{R}}=\delta_{\mathbf{\tilde{l}}\mathbf{r}} \delta_{\mathbf{\tilde{r}}\mathbf{l}}$ and $\mathcal{C}$ denotes complex conjugation in this basis. The layer inversion operator satisfies $I^2 = \mathbbm{1}$ and commutes with spin rotations and translation. It exchanges the ket-layer and bra-layer $\mathbb{Z}_2$ charges,
\begin{equation}\label{LayerInversionZ2}
I\circ (Z^{\otimes N_v}\otimes \mathbbm{1}^{\otimes N_v}) = (\mathbbm{1}^{\otimes N_v}\otimes Z^{\otimes N_v}) \circ I.
\end{equation}

\subsubsection{Hermiticity\label{sec:HermiticityTransferMatrix}}

The Hermitian conjugate of the transfer matrix is defined as $(\Gamma^{\dagger})_{\mathbf{L}\mathbf{R}}=(\Gamma_{\mathbf{R}\mathbf{L}})^*$. Since the local projection map (Eq.~\eqref{defA}) transforms in the representation $\sigma=\mathbf{A}_1+i\mathbf{A}_2$ of $C_{4v}$ the local tensor $C$ behaves under the exchange of its left and right virtual indices as $C_{ruld}=\sum_{l'r'}Y_{ll'}(C^s_{l'ur'd})^*Y_{r'r}$. Due to the $\sutwo$ symmetry of the transfer matrix this implies 
\begin{equation}\label{HermiticityTransferMatrix}
\Gamma^{\dagger}=\Gamma\circ \rhodl(e^{2\pi i S^z})
\end{equation}
such that the transfer matrix is Hermitian (anti-Hermitian) for integer (half-integer) double-layer virtual spin $\rhodl(S^z)$. Hence, the transfer matrix eigenvalues $E$ are real (purely imaginary) for integer (half-integer) spin and complex conjugation acts as $E\mapsto (-1)^{\epsilon}E$ with $\epsilon = 0$ ($\epsilon = 1$), respectively.

\subsubsection{Dressed mirror symmetry}

We denote by $\msl$ the unitary operator that reflects the virtual spins in a single-layer column about the (non-periodic) $x$-axis with basis elements
\begin{equation}
(\msl)_{\mathbf{l}\mathbf{r}}=\delta_{l_0,r_{N_v-1}}\delta_{l_1,r_{N_v-2}}\dotsm\delta_{l_{N_v-1},r_0},
\end{equation}
and by $\mdl = (\msl)^{\otimes 2}$ its double-layer variant. Transposition of the up and down virtual indices modifies the PEPS local tensor as $C_{ldru}=\sum_{d'u'}Y_{dd'}Y_{u'u}(C^s_{lu'rd'})^*$. The application of $\mdl$ to the virtual legs on the left and right of the transfer matrix therefore amounts to a complex conjugation of $\Gamma$ as well as the insertion of a matrix $Y^2=Z$ on every vertical bond in both layers. These insertions can be removed by multiplying every other horizontal virtual leg with $Z$. Hence $\Gamma$ commutes with the staggered anti-unitary operator
\begin{equation}\label{DLAntiUnitaryMirror}
\rdl \equiv \mathcal{C}\circ (Z\otimes \mathbbm{1} \otimes \dotsm \otimes Z \otimes \mathbbm{1})^{\otimes 2} \circ \mdl \circ \rhodl (e^{i\pi S^y})
\end{equation} 
where $\mathcal{C}$ denotes complex conjugation. We included the spin flip $\rhodl (e^{i\pi S^y})$ to make sure that $\rdl$ commutes with global spin rotations and that it squares to the identity $(\rdl)^2=\mathbbm{1}$. Due to its staggering the dressed mirror operator satisfies
\begin{equation}
\label{MirrorTranslation}
(\mathcal{R}_x^{dl})^{-1} \circ T^{dl} \circ \mathcal{R}_x^{dl} = (T^{dl})^{\dagger} \circ \rho_{dl}(e^{2\pi i S^z})
\end{equation}
where the last factor causes a momentum shift by $\pi$ for half-integer spin. Analogous relations hold in a single-layer column with the single-layer dressed mirror operator $\rsl \equiv \mathcal{C}\circ (Z\otimes \mathbbm{1} \otimes \dotsm \otimes Z \otimes \mathbbm{1}) \circ \msl \circ \rhosl (e^{i\pi S^y})$.

\subsection{Spectrum of $\Gamma$\label{sec:SpectrumTransferMatrix}}

The spectrum of the transfer matrix $\Gamma$ can be analyzed using the symmetries listed in Tab~\ref{tab:SymmetriesTransferMatrix}. A maximal set of Hermitian or unitary operators that commute with the transfer matrix and explains all numerically observed degeneracies is given by the spin operators, the translation operator and the $\mathbb{Z}_2$ charges. Joint eigenstates of this set are given by
\begin{equation}\label{Eigenstate}
\ket{X} = \ket{E,s_{dl},s^z_{dl},K^{dl},Z_{k},Z_{b}}
\end{equation}
where $\Gamma \ket{X} = E \ket{X}$, $\sum_{a = x,y,z}(\rhodl(S^a))^2 \ket{X} = s_{dl}(s_{dl} +1) \ket{X}$, $\rhodl(S^z) \ket{X} = s^z_{dl} \ket{X}$ and $T^{dl} \ket{X} = e^{i K^{dl}} \ket{X}$. On the other hand, the anti-unitary layer inversion and dressed mirror operators commute with the transfer matrix and spin rotations but not with the translation and the $\mathbb{Z}_2$ charges, see Eq.~\eqref{LayerInversionZ2} and Eq.~\eqref{MirrorTranslation}. Hence they create multiplets of states with the same spin quantum numbers and whose transfer matrix eigenvalues have the same absolute value $|E|$ but which have different momenta and $\mathbb{Z}_2$ charges. These multiplets are spanned by $\ket{X}$ together with the states
\begin{subequations}
	\begin{gather}
	I\ket{X} = \ket{(-1)^{\epsilon}E,s_{dl},s^z_{dl},-K^{dl},Z_{b},Z_{k}},\label{eq::inversion_halfinteger}\\
	\rdl\ket{X} = \ket{(-1)^{\epsilon}E,s_{dl},s^z_{dl},K^{dl}+\epsilon \pi,Z_{k},Z_{b}}\label{eq::dressed_halfinteger},\\
	\rdl\circ I\ket{X} = \ket{E,s_{dl},s^z_{dl},-K^{dl} + \epsilon \pi,Z_{b},Z_{k}}
	\end{gather}
\end{subequations}
where $\epsilon = 0$ ($\epsilon = 1$) and the transfer matrix eigenvalues are real (purely imaginary) for integer (half-integer) $s^{z}_{dl}$, respectively. The complex conjugation $E\mapsto (-1)^{\epsilon}E$ and the momentum inversion $K^{dl}\mapsto -K^{dl}$ are a consequence of the anti-unitarity of $I$ and $\rdl$ whereas the momentum shift $\epsilon \pi$ is caused by the spin-dependent sign factor in Eq.~\eqref{MirrorTranslation}. For half-integer spin these four states are linearly independent such that there are four redundant spectra with $Z_k\neq Z_b$.

\begin{center}
 \begin{figure}[tb]
\begin{center}
\includegraphics[width=0.95\linewidth]{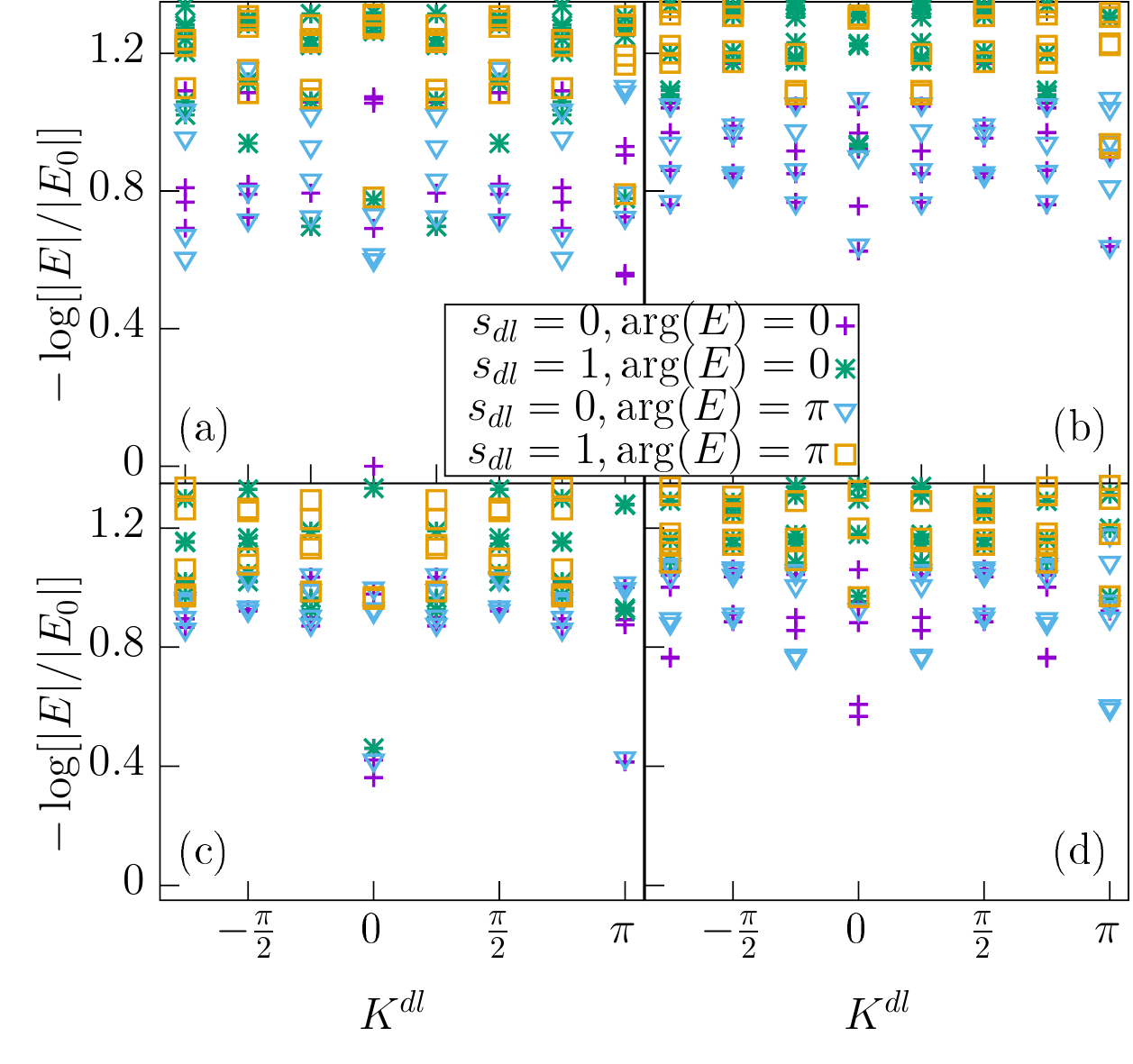}
\end{center}
\caption{Spectrum of minus the logarithm of the transfer matrix for PEPS parameters $\lambda_1 = \lambda_2 = \lambda_c = 1$, $N_v=8$ and different topological sectors $Z_{k}=Z_b =1$ in (a), $Z_{k}=Z_b=-1$ in (b), $Z_{k}=Z_b =1$ and a vison line in both layers in (c), and $Z_{k}=Z_b =-1$ and a vison line in both layers in (d). These spectra are expected to give the minima of the dispersion relation of the topologically trivial excitations of a local parent Hamiltonian above the corresponding topological ground state at the point $k_x = \text{arg}(E), k_y = K^{dl}$ in the two-dimensional Brillouin zone (see main text). In all cases we show only the first few leading eigenvalues per spin and momentum sector. States with higher spin appear at higher values of $-\log[|E|/|E_0|]$ and are not shown here.}
\label{fig::Spec_Gamma_SZ_0} 
\end{figure}
\end{center}

On the other hand, for integer spin the dressed mirror symmetry does not cause any degeneracies~\footnote{Since $(\rdl)^2=\mathbbm{1}$ and $I^2=\mathbbm{1}$ Kramer's theorem for anti-unitary operators does not apply such that the mirror symmetry and layer inversion do not cause any degeneracies whenever they do not change any quantum numbers.}. The two-dimensional multiplet for $K^{dl}\neq 0,\pi$ is spanned by $\ket{X},I\ket{X}$ whereas these two states are identical for $K^{dl}= 0,\pi$~\cite{Note1}. The spectrum of minus the logarithm of the transfer matrix without flux lines for $\lambda_1 = \lambda_2 = \lambda_c = 1$ and $N_v=8$ is shown in Fig.~\ref{fig::Spec_Gamma_SZ_0}(a) for $Z_{k}=Z_b =1$ and Fig.~\ref{fig::Spec_Gamma_SZ_0}(b) for $Z_{k}=Z_b =-1$. This spectrum is expected to reproduce the relative energy of the minima of the dispersion relation of a local parent Hamiltonian at the point $k_x = \text{arg}(E), k_y =K^{dl}$ in the two-dimensional Brillouin zone (BZ)~\cite{Zauner2015,haegeman2015shadows}. Due to the reflection symmetry discussed in Sec.~\ref{sec:HermiticityTransferMatrix} the transfer matrix in these sectors is Hermitian with real eigenvalues such that we probe two lines of the BZ where $k_x =0,\pi$. Since each plot contains only states in the same topological sector of the $\mathbb{Z}_2$ symmetry the data we show describes topologically trivial excitations which are expected to be identical in every topological sector in the thermodynamic limit. Indeed, the dispersion relation for states with given spin is similar in all considered topological sectors with minima at $K^{dl}=0,\pi$ and maxima at $K^{dl}=\pi/2,3\pi/2$. The spectrum appears gapless without visible branches of single-particle excitations in agreement with the expectation that chiral topological PEPS are gapless~\cite{wahl2013projected,DubailPRB2015}. An exception is the lowest level at zero momentum in the $\mathbb{Z}_2$-even sector whose gap to the rest of the states may be a finite-size effect~\cite{DubailPRB2015}. The quasi-energy of the lowest excitations with given spin grows with $s_{dl}$ such that the leading states all have zero spin. For $N_v=8$, the lowest levels in either topological sector lie at zero momentum and we will use the corresponding states to compute the ES in the next section.

\subsection{Transfer matrix $\Gammah$ with horizontal flux string\label{sec:TransferMatrix_string}}

The ground state manifold for a PEPS with a virtual $\mathbb{Z}_2$ symmetry includes the state $\ket{\psi}_{(h)}$ with a horizontal flux string along the cylinder as described in section~\ref{sec:VirtualSymmetry}. The transfer matrix $\Gammah$ for this state is obtained from $\Gamma$ by the insertion of a matrix $Z$ on the vertical virtual bond between the sites $(N_v-1,0)$ both in the ket- and the bra-layer.

$\Gammah$ possesses the same symmetries as $\Gamma$ except for translation which is replaced by a dressed translation operator which accounts for the changing position of the flux line. Indeed, translation by one lattice site in the vertical direction causes a shift of the $Z$-insertions from the bond $(N_v-1,0)$ to the bond $(0,1)$. The insertions can be returned to their original position by multiplying the external virtual legs of site $0$ in both layers with matrices $Z$ and exploiting the virtual $\mathbb{Z}_2$ symmetry of the local tensors on that site. The transfer matrix with a flux line therefore commutes with a unitary dressed translation operator
\begin{equation}
\tddl = (\tdsl) ^{\otimes 2} = ( T^{sl} \circ (Z\otimes \mathbbm{1} \otimes \mathbbm{1} \otimes \dotsm \otimes \mathbbm{1}))^{\otimes 2}.
\end{equation}In contrast to the usual translation operator, the dressed translation satisfies
\begin{equation}
(\tddl)^ {N_v}= \rhodl(e^{2\pi i S^z})
\end{equation}and thereby imposes anti-periodic boundary conditions for half-integer virtual spins. Correspondingly, the momentum takes integer (half-integer) values in the sector of integer (half-integer) spin $\rhodl(S^z)$. Analogous statements hold for the single-layer dressed translation operator and momentum. The dressed translation operator satisfies Eq.~\eqref{MirrorTranslation} with the anti-unitary dressed mirror operator such that Table~\ref{tab:SymmetriesTransferMatrix} applies also to $\Gammah$ after substitution of $\tddl$ for $T^{dl}$.  

The spectrum of minus the logarithm of the transfer matrix with a horizontal flux line in both layers for $\lambda_1 = \lambda_2 = \lambda_c = 1$ and $N_v=8$ is shown in Fig.~\ref{fig::Spec_Gamma_SZ_0}(c) for $Z_{k}=Z_b =1$ and in Fig.~\ref{fig::Spec_Gamma_SZ_0}(d) for $Z_{k}=Z_b =-1$. In the $\mathbb{Z}_2$-odd sector the spectrum is very similar to the case without flux line. In the $\mathbb{Z}_2$-even sector the dispersion of the states in the continuum appears flatter than without flux line. The minima at $K^{dl}=0,\pi$ split into multiple levels each and appear at almost the same quasi-energy that is significantly above the lowest state without flux line. Again, the leading states in either topological sector that we will use to compute the ES lie at zero spin and momentum.

\subsection{Entanglement spectrum\label{sec:SymmetriesES}}

In this section we investigate the ES of the chiral PEPS on an infinite cylinder in all topological sectors of the $\mathbb{Z}_2$ symmetry. As explained in Secs.~\ref{sec:ESPEPS} and~\ref{sec:TransferMatrix} the ES is given by the leading eigenvectors of the transfer matrix according to Eq.~\eqref{ESHermitianDensityMatrix}. We begin by analyzing the symmetries of the ES, thereby providing an analytical explanation for the observation of two branches in the ES corresponding to the spin-$\frac{1}{2}$ field of $\mathfrak{su}(2)_1$~\cite{Poilblanc2015PRB_chiralPEPS,Poilblanc2016PRB_chiralPEPS}.

\subsubsection{Symmetries of the entanglement spectrum\label{sec:SymmetriesESAnalytical}}

Any eigenstate $\ket{X}$ of the transfer matrix  defines a virtual reduced density matrix $\sigma_X$ with matrix elements $(\sigma_X)_{\mathbf{l}\mathbf{\tilde{l}}}= X_{(\mathbf{l},\mathbf{\tilde{l}})}$. If $\ket{X}$ is an eigenvector with eigenvalue $\mu$ of a double-layer operator $\mathcal{O}_{1}\otimes \mathcal{O}_2$ the associated $\sigma_X$ satisfies the relation
\begin{equation}\label{CommutationSL}
\mathcal{O}_{1} \circ \sigma_X \circ \mathcal{O}_{2}^T = \mu \,\sigma_X
\end{equation}
where $\mathcal{O}_{1,2}$ are single-layer operators. Applied to the double-layer spin (translation)  Eq.~\eqref{CommutationSL} implies that the virtual density matrix of a double-layer eigenstate with $s_{dl}=0$ ($K^{dl}=0$) commutes with the single-layer spin (translation). On the other hand, $Z_{k,b}$ determine the eigenvalue of $Z^{\otimes N_v}$ on the image and support of $\sigma_X$. Eq.~\eqref{CommutationSL} applies also to anti-unitary double-layer operators such as the mirror $\rdl$. Its fixed points are therefore associated with mirror-symmetric reduced density matrices $\sigma_X\rsl=Z^{\otimes N_v}\rsl \sigma_X$. Moreover, layer inversion maps the virtual density matrix to its Hermitian conjugate such that fixed points of $I$ correspond to Hermitian virtual density matrices.

We now show that the ES derived from a transfer matrix eigenstate $\ket{X}$ with zero spin and momentum and $Z_k =Z_b = -1$ contains two degenerate branches related by a momentum shift $K^{sl}\mapsto K^{sl} + \pi$. Provided that there are no accidental degeneracies, we can choose the phase of $\ket{X}$ such that the density matrix $\sigma_X$ is Hermitian and commutes or anti-commutes with $\rsl$. In the $\mathbb{Z}_2$-odd sector the dressed mirror causes a momentum shift of $\pi$ according to Eq.~\eqref{MirrorTranslation}. Hence, any density matrix eigenstate $\sigma_x\ket{v} = e^{-\xi/2}\ket{v}$ has an orthogonal mirror image $\rsl\ket{v}$ with the same entanglement energy $\xi$ but momentum shifted by $\pi$. As long as the dressed mirror symmetry is not broken, the ES of this state therefore contains two exactly degenerate branches shifted in momentum by $\pi$.

\subsubsection{Numerical results}

\begin{center}
 \begin{figure}[htb]
\begin{center}
\includegraphics[width=0.99\linewidth]{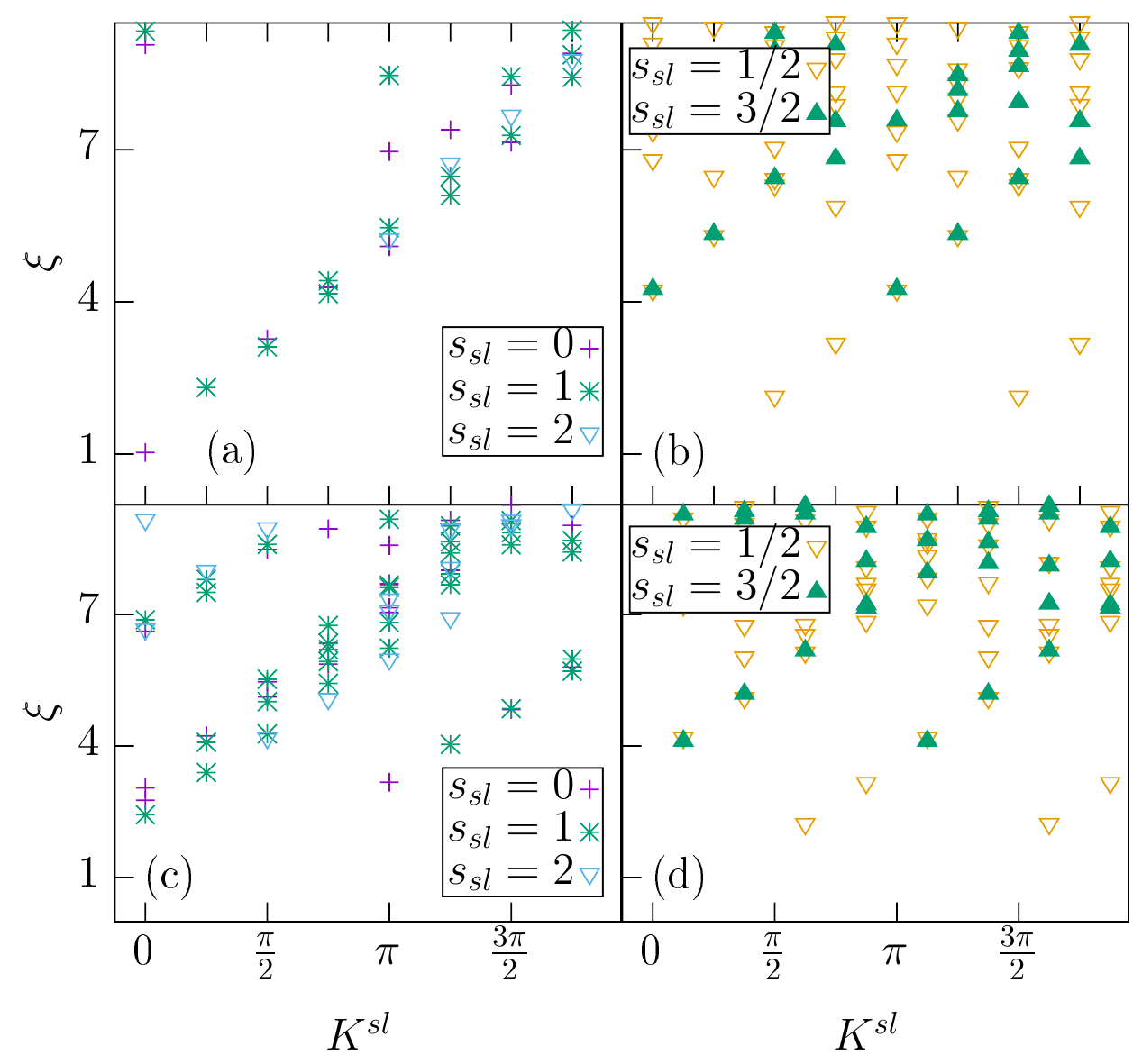}
\end{center}
\caption{Entanglement spectrum for the leading eigenvectors of the transfer matrix with $\lambda_1 = \lambda_2 = \lambda_c = 1$ and $N_v=8$ in the diagonal sectors with $Z_{k}=Z_{b} = 1$ in (a), (c) and $Z_{k}=Z_{b} = -1$ in (b), (d) and a vison line in both layers for (c), (d). The ES in (a) has a single chiral mode with the integer-spin state counting of the chiral CFT $\mathfrak{su}(2)_1$. The ES in (b) and (d) are very similar and are symmetric under a momentum shift of $\pi$ caused by the dressed mirror operator. In the region $0\leq K^{sl} \leq \pi$ they have a single chiral mode with the half-integer-spin state counting of $\mathfrak{su}(2)_1$. The ES in (c) has some chiral features but possesses multiple branches whose state counting could not be related to a simple CFT.}
\label{fig::ES} 
\end{figure}
\end{center}

The ES of the leading eigenstates is shown in Fig.~\ref{fig::ES} for $\lambda_1 = \lambda_2 = \lambda_c = 1$ and $N_v=8$. The ES for the PEPS without flux line in (a) and (b) display a linear dispersion relation where the number of low-lying states per momentum sector is that expected for the chiral CFT $\mathfrak{su}(2)_1$. Due to the dressed mirror symmetry of the transfer matrix the spectrum (b) in the half-integer spin sector possesses two branches shifted in momentum by $\pi$ as discussed above. Moreover, the estimated conformal weight of the half-integer sector is much larger than the expected value of $1/4$. In the half-integer spin sector the ES in Fig.~\ref{fig::ES}(d) with a flux line looks very similar to the ES without flux line up to an overall momentum shift of $1/2$ due to the anti-periodic boundary conditions. However, the ES with a horizontal flux line in the integer spin sector has some chiral features but does not possess a single mode with the state counting of $\mathfrak{su}(2)_1$. Up to now, we were not able to find a simple CFT with the state counting of this ES.

\section{Chiral spin liquid PEPS for $\lambda_1=0$\label{sec:ZeroLambdaOne}}

In this section we study the chiral spin-$\frac{1}{2}$ liquid PEPS for $\lambda_1 = 0$. This case is interesting since the PEPS possesses an additional virtual $\uone$ symmetry provided that the lattice is bipartite. Such a symmetry appears also for the nearest-neighbour RVB PEPS~\cite{Poilblanc2012PRB_RVB} albeit with a different charge per unit cell. After defining the $\uone$ symmetry in Sec.~\ref{sec:UOneSymmetry} we analyse its implications for the transfer matrix spectrum in Sec.~\ref{sec:SpectrumTransferMatrixUOne}. Moreover we provide numerical evidence that not all charge sectors lead to independent physical states and identify the dominant sectors. In Sec.~\ref{sec:ESUOne} we compute the ES of the leading independent states with fixed charge and show that the half-integer spin sector has a single chiral branch and does not contain any degeneracies created by the dressed mirror symmetry. Furthermore we show that the estimated conformal weight is very close to the value expected for $\mathfrak{su}(2)_1$ when $\lambda_c\approx \lambda_2$.

\subsection{Staggered virtual $\uone$ symmetry\label{sec:UOneSymmetry}}

The virtual Hilbert space $\mathbf{0}\oplus\mathbf{\frac{1}{2}}$ carries a virtual $\uone$ action $U({\varphi})$ defined in Eq.~\eqref{UOneAction} that changes the relative phase of vectors in the two spin subspaces. Generically this phase rotation maps the local PEPS tensor to a different point in the parameter space, $(\lambda_1,\lambda_2,\lambda_c)\mapsto(\lambda_1,e^{2i\varphi}\lambda_2,e^{2i\varphi}\lambda_c)$ (see Eq.~\eqref{UOneTensor}). However, if either $\lambda_2=\lambda_c=0$ or $\lambda_1 = 0$ all non-vanishing configurations in the local tensor acquire the same phase under this operation such that
\begin{equation}\label{UOneTensorSymmetric}
\amap \circ U(\varphi)^{\otimes 4} = e^{in\varphi} \amap,
\end{equation} 
where the charge is $n = 3$ ($n = 1$) for $\lambda_1 = 0$ ($\lambda_2=\lambda_c=0$). This identity is very similar to the definition of the virtual $\mathbb{Z}_2$ symmetry in Eq.~\eqref{VirtualZ2Symmetry}. However, Eq.~\eqref{UOneTensorSymmetric} implies the existence of a global virtual $\uone$ symmetry for the PEPS only if the lattice is bipartite. Indeed, invariance of the state requires that the left and right (top and bottom) virtual legs of every unit cell transform in mutually conjugate representations. On a bipartite lattice with a unit cell of $2\times 2$ sites one can therefore construct a staggered symmetry by combining transformations $U(\varphi)$  and $U(-\varphi)$ for sites on $\Lambda_A$ and $\Lambda_B$, respectively. This staggering of the $\uone$ symmetry will become crucial in order to obtain an ES with only a single branch in the half-integer sector.

When expressed in terms of the generator
\begin{equation}\label{UOneGenerator}
Q=-i(\partial_{\varphi} U(\varphi))|_{\varphi = 0} = \begin{pmatrix}
0 & 0 & 0 \\
0 & 1 & 0 \\
0 & 0 & 1
\end{pmatrix}
\end{equation}
which counts the number of virtual legs in the spin-$\frac{1}{2}$ virtual representation, the relation~\eqref{UOneTensorSymmetric} translates into a virtual Gauss law
\begin{multline}\label{GaussLaw}
\amap \,\circ \big[Q\otimes \mathbbm{1}\otimes \mathbbm{1}\otimes \mathbbm{1} + \mathbbm{1}\otimes Q\otimes \mathbbm{1}\otimes \mathbbm{1} + \\
\mathbbm{1}\otimes \mathbbm{1}\otimes Q\otimes \mathbbm{1}+\mathbbm{1}\otimes \mathbbm{1}\otimes \mathbbm{1}\otimes Q\big] = n \,\amap.
\end{multline}
For $\lambda_1=0$, $n=3$ in Eq.~\eqref{GaussLaw} implying that exactly three virtual legs of every local tensor are in the spin-$\frac{1}{2}$ state. On the other hand, for the nearest-neighbour RVB state obtained at $\lambda_2=\lambda_c=0$ one finds $n=1$ such that exactly one virtual leg is in the spin-$\frac{1}{2}$ state~\cite{Poilblanc2012PRB_RVB}. 

Similarly to the case of discrete virtual symmetries one can consider states obtained from the chiral PEPS by inserting strings of virtual $\uone$ operators along non-contractible loops of the manifold. For any local parent Hamiltonians these states are indistinguishable from the original PEPS when $\lambda_1 =0$. On a cylinder the minimally entangled states with respect to a vertical entanglement cut are generated by insertion of flux strings in the horizontal direction on one hand and projection on sectors of fixed charge in the vertical direction on the other hand. The conserved charge of the staggered virtual $\uone$ symmetry acting on one column of virtual boundary legs is
\begin{equation}\label{UOneGeneratorSingleLayer}
Q_{sl} = \sum_{i = 0}^{N_v-1} (-1)^i Q_{(i)}
\end{equation}
with integer eigenvalues $q$ in the range $-N_v/2 \leq q \leq N_v/2$. In Eq.~\eqref{UOneGeneratorSingleLayer}, the subscript $i$ indicates on which virtual spin the local generator~\eqref{UOneGenerator} acts and the factor $(-1)^i$ accounts for the staggering of the virtual symmetry. The local $\mathbb{Z}_2$ matrix is a special case of a $\uone$ rotation, $Z = U(\pi)$. Therefore, even (odd) charges $q$ correspond to trivial (non-trivial) $\mathbb{Z}_2$ charge. Moreover, the single-layer spin in a given charge sector is constrained as
\begin{equation}\label{BoundUOneSpin}
|2s_{sl}^z |\leq N_v - |q|.
\end{equation}
In the following we focus on analyzing the PEPS with fixed vertical charge $q$ and without horizontal $\uone$ flux strings. 

\subsection{Transfer matrix\label{sec:SpectrumTransferMatrixUOne}}

In this section we analyze the consequences of the staggered virtual $\uone$ symmetry at $\lambda_1=0$ for the spectrum of the transfer matrix $\Gamma$ on a cylinder of even $N_v$.

\subsubsection{Quantum numbers}

Due to the virtual Gauss law~\eqref{GaussLaw} the single-column transfer matrix anti-commutes with the generator~\eqref{UOneGeneratorSingleLayer} applied to both the ket- and bra-layer separately,
\begin{subequations}
\label{UOneGeneratorDoubleLayer}
\begin{gather}
Q_k = Q_{sl}\otimes \mathbbm{1},\\
Q_b =\mathbbm{1}\otimes Q_{sl}.
\end{gather}
\end{subequations} 
As discussed above, the corresponding charges $q_k$ and $q_b$ in the ket- and bra-layer take values $-N_v/2\leq q_k,q_b\leq N_v/2$ and define a bound for the double-layer spin,
\begin{equation}\label{BoundUOneSpinDoubleLayer}
|2s_{dl}^z| \leq \big|2N_v - |q_k| - |q_b|\big|.
\end{equation}
The double-column transfer matrix $\Gamma^2$ for the PEPS on a patch of size $2\times N_v$ therefore commutes with $Q_{k,b}$ and thus possesses a $\uone\times \uone$ symmetry with generators~\eqref{UOneGeneratorDoubleLayer}.

It is natural to analyze the spectrum of the transfer matrix in terms of the $\uone$ charges $q_k,q_b$. However, this is not compatible with the set of quantum numbers from Tab.~\ref{tab:SymmetriesTransferMatrix} that we used for the single-column transfer matrix at arbitrary PEPS parameters $\lambda_1,\lambda_2,\lambda_c$. Indeed, due to their staggering the generators $Q_{k,b}$ generally anti-commute both with the translation operator and the single-column transfer matrix. We therefore pass to a modified set of quantum numbers given by the double-column transfer matrix $\Gamma^2$, the $\uone$ generators $Q_k,Q_b$, the $\sutwo$ spin operators, double-step translation $(T^{dl})^2$ and the product $\Gamma T^{dl}$ of the single-column transfer matrix and single-step translation (individually, both of these operators anti-commute with the $\uone$ generators). We refer to joint eigenstates of this commuting set as 
\begin{equation}\label{Eigenstate2}
\ket{\tilde{X}} = \ket{E^2, s_{dl}, s^z_{dl}, \tilde{K}^{dl}, \mu, q_k, q_b},
\end{equation}
where $\Gamma^2\ket{\tilde{X}} = E^2 \ket{\tilde{X}}$, the spin quantum numbers are as defined above and the momentum of the double-step translation takes values $\tilde{K}^{dl} = 2\pi n /(N_v/2)$ with $0\leq n \leq N_v/2 -1$. The quantum number $\mu$ is defined as the eigenvalue of $\Gamma T^{dl}$ and satisfies $\mu^2 = E^2 e^{i \tilde{K}^{dl}}$ so that it can take only two values once the transfer matrix weight and double-step momentum are fixed.

Alternatively one can use the set of quantum numbers from Tab.~\ref{tab:SymmetriesTransferMatrix} extended by the products $q_k^2, q_b^2, q_kq_b$ which are compatible with single-step translation. The resulting joint eigenstates
\begin{equation}\label{Eigenstate1}
\ket{X} = \ket{E, s_{dl}, s^z_{dl}, K^{dl}, q_k^2, q_b^2, q_kq_b}
\end{equation}   
generally are linear superpositions of multiple eigenstates $\ket{\tilde{X}}$ defined in Eq.~\eqref{Eigenstate2}.

In the following we will use the basis given by Eq.~\ref{Eigenstate2} to analyze the transfer matrix spectrum. A special situation arises in the sector with $q_k = q_b =0$ where $Q_{k,b}$ are identically zero and therefore commute with the single-column transfer matrix and single-step translation operator. In this case we recover the single-step momentum as a quantum number and the eigenstates Eq.~\eqref{Eigenstate1} are in one-to-one correspondence with the states Eq.~\eqref{Eigenstate2}.

\subsubsection{Multiplets}

The spectrum of the square of the transfer matrix contains multiplets of states with the same weight $E^2$ and spin quantum numbers but different values for the momentum, $\uone$ charges or quantum number $\mu$. In terms of the eigenstates defined in Eq.~\eqref{Eigenstate2} these multiplets are created by layer inversion $I$, the dressed mirror symmetry $\rdl$ and their combination which act on the quantum numbers as~\footnote{For system sizes $N_v \in 4\mathbb{N}$ the quantum number $\mu$ permits us to distinguish the two points $\tilde{K}^{dl} = 0$ and $\tilde{K}^{dl} = \pi$ which are both invariant under inversion of the double-step momentum $\tilde{K}^{dl}$ whereas only $\tilde{K}^{dl} = 0$ corresponds to inversion-symmetric single-step momenta $K^{dl} = 0, \pi$. Indeed, $\mu$ is real for $\tilde{K}^{dl} = 0$ and purely imaginary for $\tilde{K}^{dl} = \pi$.}
\begin{subequations}
	\begin{align}
	I: &\quad \tilde{K}^{dl}\mapsto -\tilde{K}^{dl},\, \mu \mapsto \mu^*, \,q_k \leftrightarrow q_b\\
	\rdl: &\quad \tilde{K}^{dl}\mapsto \tilde{K}^{dl}, \,\mu \mapsto \mu, \,q_{k,b} \mapsto -q_{k,b}.
	\end{align}
\end{subequations}
Hence there are four degenerate states whenever (i)~$|q_k| \neq |q_b|$ or (ii)~$|q_k| = |q_b| \neq 0$ and $\tilde{K}^{dl}\neq 0$. Similarly one finds that there are two degenerate states when (iii)~$|q_k| = |q_b| \neq 0$ and $\tilde{K}^{dl}=0$ or (iv)~$q_k = q_b= 0 $ and $\tilde{K}^{dl}\neq 0$. Compared to general values of the parameters $(\lambda_1,\lambda_2,\lambda_c)$ the spectrum at $\lambda_1 = 0$ therefore contains larger multiplets of degenerate states.

For any given diagonal sector with non-zero charge $q_k= q_b = q$ this discussion implies that the leading transfer matrix eigenstate
\begin{equation}\label{EigenstateUOne1}
\ket{\tilde{X}_0}_{(q,q)}
\end{equation}
is exactly degenerate with the leading state with opposite charge $q_k= q_b = -q$. Numerically we found that these leading states have zero spin as well as zero momentum with respect to the double-step translation operator. Therefore they are invariant under layer inversion $I$ and their degeneracy is due to the dressed reflection symmetry of the transfer matrix. Numerically we observe that there are no additional degeneracies. Since single-step translation also reverses the sign of the charges, $q_{k,b} \mapsto -q_{k,b}$, we can therefore choose a basis such that
\begin{equation}\label{EigenstateUOne2}
\ket{\tilde{X}_0}_{(-q,-q)} = T^{dl} \,\ket{\tilde{X}_0}_{(q,q)}.
\end{equation} 
Thus we can construct eigenstates with well-defined single-step momentum $K^{dl} = 0,\pi$ and $q_k^2= q_b^2= q_kq_b = q^2$ as linear superpositions
\begin{equation}\label{EigenstateMomentum}
\ket{X_0}_{K^{dl} = 0,N} = \frac{1}{\sqrt{2}}[\ket{\tilde{X}_0}_{(q,q)} \pm \ket{\tilde{X}_0}_{(-q,-q)}]
\end{equation}of the two degenerate states with well-defined charges. The states~\eqref{EigenstateUOne1} and \eqref{EigenstateUOne2} are not stable under perturbations of the PEPS induced by small non-vanishing values of $\lambda_1$ due to the breaking of the $\uone$ symmetry for any non-vanishing $\lambda_1$. On the other hand we expect that to first order in $\lambda_1$ the states~\eqref{EigenstateMomentum} remain eigenstates of the transfer matrix although degeneracy will be lifted.

\subsubsection{Dominant sectors on finite-site cylinders}

The chiral PEPS at $\lambda_1 =0$ gives rise to an extensive number of states obtained by projecting the virtual boundary vector at one end of the cylinder onto a given eigenvalue sector of the conserved charge $Q_{sl}$. However, it is not clear whether all of these boundary sectors correspond to independent physical states in the thermodynamic limit. Indeed, some of these states may have vanishing norm or correspond to a linear superposition of other states in the limit $N_v\rightarrow\infty$ as has been observed for free fermionic chiral PEPS~\cite{Wahl2014PRB}. For long cylinders the quotient of the norm of two states $\ket{\psi_q}$ and $\ket{\psi_p}$ with fixed boundary charges $q$ and $p$ is expected to approach~\cite{Schuch2013PRL}
\begin{equation}\label{Norm}
\langle \psi_q | \psi_q\rangle / \langle \psi_p | \psi_p\rangle \approx \Big(\frac{E_{q,q}}{E_{p,p}}\Big)^{N_h/2}
\end{equation}
where $E_{q,q}$ refers to the dominant eigenvalue of the double-column transfer matrix in the diagonal sector where the bra-layer and ket-layer charges are $q_b = q_k = q$. Moreover, the dominant eigenvalues of the transfer matrix in the off-diagonal sectors with $q_k\neq q_b$ determine the overlap of two normalised states~\cite{Schuch2013PRL}
\begin{equation}\label{Overlap}
\langle \psi_q | \psi_p\rangle \approx \Big(\frac{E_{q,p}}{\sqrt{E_{q,q}E_{p,p}}}\Big)^{N_h/2}.
\end{equation}

\begin{center}
 \begin{figure}[tbh]
\begin{center}
\includegraphics[width=0.95\linewidth]{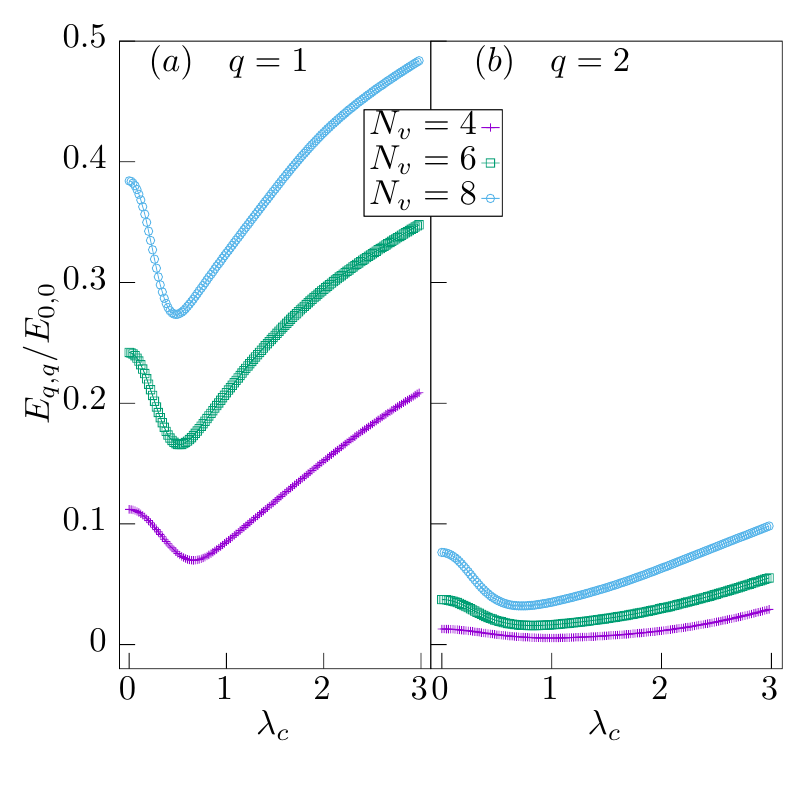}
\end{center}
\caption{Normalised leading eigenvalue $E_{q,q}/E_{0,0}$ of the double column transfer matrix in the diagonal sector $q_k = q_b = q$ as a function of $\lambda_c$ for $\lambda_1=0$, $\lambda_2 = 1$, different cylinder widths $N_v$ and $q=1$ in (a) and $q=2$ in (b). On long cylinders, this ratio is expected to determine the relative norm of the states in different $\uone$ sectors according to Eq.~\eqref{Norm}.} 
\label{fig::ScalingNorm} 
\end{figure}
\end{center}

\begin{center}
 \begin{figure}[tbh]
\begin{center}
\includegraphics[width=0.95\linewidth]{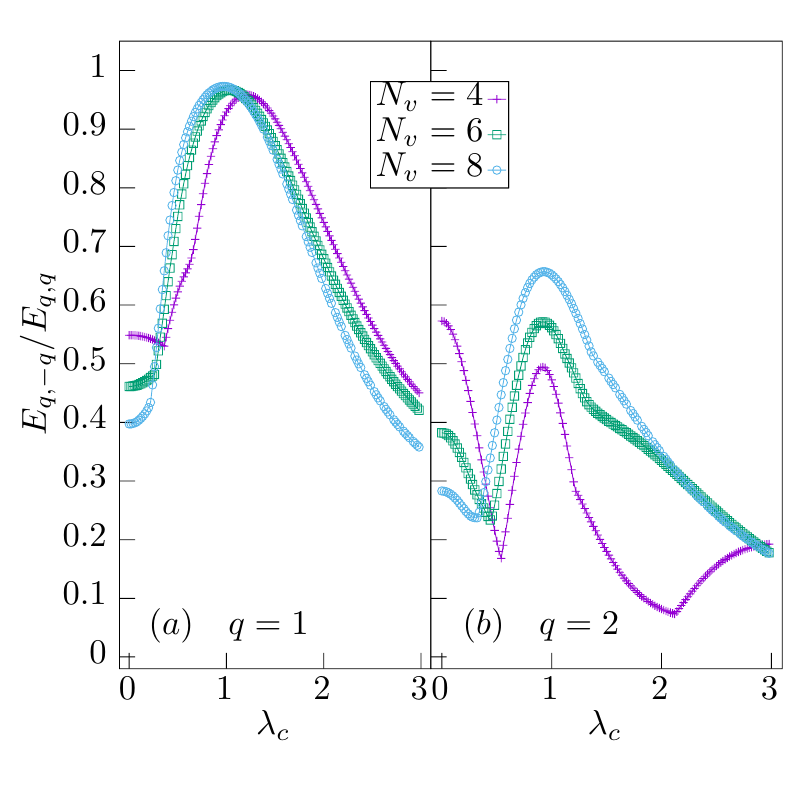}
\end{center}
\caption{Ratio of the leading transfer matrix eigenvalues in the off-diagonal and diagonal sectors with $q_k = - q_b = q$ and $q_k = q_b = q$ as a function of $\lambda_c$ for $\lambda_1=0$, $\lambda_2 = 1$, different cylinder widths $N_v$ and $q=1$ in (a) and $q=2$ in (b). On long cylinders, this ratio is expected to determine the overlap of the states in different $\uone$ sectors according to Eq.~\eqref{Overlap}. Since the normalised overlap for $q=1$ is very close to unity when $\lambda_c\approx\lambda_2$ we expect that the states $\ket{\psi_1}$ and $\ket{\psi_{-1}}$ become identical in the thermodynamic limit in this region.}
\label{fig::ScalingOverlap} 
\end{figure}
\end{center}

In order to study the overlap and weight of different sectors we have performed exact diagonalization of the double-column cylinder transfer matrix including the quantum numbers $q_k, q_b$ for the numerically accessible values of the cylinder width $N_v$. The normalized leading eigenvalue $E_{q,q}/E_{0,0}$ in the diagonal sector for $\uone$ charges $q=1,2$ is displayed in Fig.~\ref{fig::ScalingNorm} as a function of the PEPS parameter $\lambda_c$ for $\lambda_2=1$ and for system sizes $N_v = 4,6,8$. For all considered values of $\lambda_c$ and $N_v$, the ratio $E_{q,q}/E_{0,0}$ decreases with increasing $q$ and is much smaller than unity if $|q| > 1$ (for $N_v = 8$ we find that $E_{3,3}/E_{0,0} \sim 10^{-3}$ and $E_{4,4}/E_{0,0} \sim 10^{-4}$, respectively). The normalized overlap $E_{q,-q}/E_{q,q}$ for $q=1,2$ and $N_v = 4,6,8$ is displayed as a function of $\lambda_c$ in Fig.~\ref{fig::ScalingOverlap}. For both values of $q$ and all system sizes the overlap increases rapidly for $0.5\leq\lambda_c\leq 1$, has a maximum in the vicinity of $\lambda_c = 1$ and decreases again for bigger values of $\lambda_c$. The overlap for $q=1$ ($q=2$) appears to increase with $N_v$ for $\lambda_c \leq 1.3$ ($\lambda_c \leq 2$). For $N_v =8$ the maximal overlap for $q=1$ is above $97 \%$ whereas for $q=2$ the maximal value is around $65\%$ but does not appear to be converged as a function of the system size. All in all, the dominant sectors for small cylinder width $N_v$ are therefore those with small $\uone$ charges $|q| \leq 1$ where the states with $q=1$ and $q=-1$ that are related by a single-step translation have a very large overlap above $97\%$. Since we are restricted to small systems the finite-size scaling in Figs.~\ref{fig::ScalingNorm} and \ref{fig::ScalingOverlap} is not conclusive. However, it is consistent with the hypothesis that the only two independent leading sectors in the thermodynamic limit are those with $q=0$ and $q=1$.  

\subsection{Entanglement spectrum}\label{sec:ESUOne}

The ES for a PEPS on an infinite cylinder is obtained from all transfer matrix eigenstates whose eigenvalues approach the leading eigenvalue in the limit $N_v\rightarrow\infty$. Here, the contributing states should form an orthogonal set. If the PEPS has a virtual symmetry the correct relative weight for the leading states in different symmetry sectors in the ES is \textit{a priori} not known and should be chosen such that the entanglement Hamiltonian as an operator for the one-dimensional spin chain at the virtual cut is as local as possible~\cite{Schuch2013PRL}. As discussed in the previous subsection the leading sectors at finite cylinder width for the chiral PEPS at $\lambda_1=0$ are those with conserved $\uone$ charge $q=-1,0,1$. In particular we exclude sectors with charge $|q|\geq 2$ since their transfer matrix eigenvalues are far suppressed. Moreover we include only one of the states with $q=\pm 1$ since we expect that they become identical in the thermodynamic limit. The virtual reduced density matrix is then a linear superposition of the density matrices $\sigma_{q}$ for the leading states in the sectors $q=0,1$ which are individually normalized to $\Tr \sigma_q^2 =1$. The relative weight in this linear superposition determines the overall shift in entanglement energy between levels with integer and half-integer spin and thereby also the conformal weight of the spin-1/2 field. We expect that the ansatz
\begin{equation}\label{SigmaLinSup}
\sigma = \sigma_0 + \sigma_1
\end{equation}
for the virtual reduced density matrix correctly reproduces the low-lying levels in the ES. Indeed, the linear superposition~\eqref{SigmaLinSup} possesses equal weight on the $\mathbb{Z}_2$ even and odd sectors as expected for a PEPS with a virtual $\mathbb{Z}_2$ symmetry~\cite{Schuch2013PRL}.

In the following subsection (Sec.~\ref{sec:ESUOneAnalytical}) we further motivate the ansatz Eq.~\eqref{SigmaLinSup} for the virtual reduced density matrix by showing analytically that the two density matrices $\sigma_{\pm q}$ possess an identical ES without any degeneracies caused by the dressed mirror symmetry. On the other hand the ES of the translation invariant linear superposition Eq.~\eqref{EigenstateMomentum} contains two copies of the ES of $\sigma_{\pm q}$ shifted in momentum by $\pi$. Finally, in Sec.~\ref{sec:ConformalWeight} we discuss the chirality of the ES derived from Eq.~\eqref{SigmaLinSup} for $N_v=8$ and show that the conformal weight of the spin-$1/2$ sector is very close to the expected value of $1/4$ when $\lambda_c\approx\lambda_2$.

\subsubsection{ES for $q\neq 0$}\label{sec:ESUOneAnalytical}

For any value of the $\uone$ charge $q$, the transfer matrix eigenstate $\ket{\tilde{X}_0}_{(q,q)}$ corresponds to a virtual reduced density matrix $\sigma_q$ that maps the virtual spins in the bra-layer of the entanglement cut to those in the ket-layer as explained in Sec.~\ref{sec:SymmetriesESAnalytical}. Due to the fixed charge the support and image of $\sigma_{q}$ consist of the subspace of the single-layer virtual space with $Q_{sl} = q$ such that
\begin{equation}
Q_{sl}\, \sigma_{q} = \sigma_{q}\, Q_{sl} = q \, \sigma_{q}.
\end{equation} 
Since $\ket{\tilde{X}_0}_{(q,q)}$ is invariant under double-step translation, the virtual reduced density matrix $\sigma_{q}$ commutes with the single-layer double-step translation $(T^{sl})^2$. 

Whenever the leading states $\ket{\tilde{X}_0}_{(\pm q,\pm q)}$ with opposite non-vanishing charges are related by single-step translation as in Eq.~\eqref{EigenstateUOne2} their virtual reduced density matrices are unitarily equivalent with the basis change given by single-layer single-step translation,
\begin{equation}
\sigma_q = T^{sl} \, \sigma_{-q} \, (T^{sl})^{\dagger}.
\end{equation}
Hence the ES of $\ket{\tilde{X}_0}_{(\pm q,\pm q)}$ given by the spectra of $-\log \sigma_{\pm q}^2$ are identical. This ES does not possess any degeneracies caused by the anti-unitary dressed mirror symmetry since $\rsl$ does not act within a subspace of fixed non-zero $\uone$ charge $q$.

On the other hand the virtual reduced density matrix of the momentum eigenstate $\ket{X_0}_{K^{dl} = 0}$ defined in Eq.~\eqref{EigenstateMomentum} is given by the sum
\begin{equation}\label{eq:newdensitymatrix}
\sigma_{+} = \frac{1}{\sqrt{2}}[\sigma_q + \sigma_{-q}]
\end{equation}
which has support on both the $Q_{sl} = \pm q$ subspaces. Eigenstates of $\sigma_+$ with well-defined single-step momentum $K^{sl}$ are given by linear superpositions of eigenstates of $\sigma_{\pm q}$ with well-defined double-step momentum $\tilde{K}^{sl}$. For a relative phase of $\pm1$ in the linear superposition one finds either $K^{sl} = \tilde{K}^{sl}$ or $K^{sl} = \tilde{K}^{sl}+ \pi$. Therefore the ES of $\ket{X_0}_{K^{dl} = 0}$ consists of two copies of the ES of $\ket{\tilde{X}_0}_{(\pm q,\pm q)}$ shifted in momentum by $\pi$. Note that the normalized ES of $\ket{X_0}_{K^{dl} = 0}$ is shifted by $\ln 2$ compared to that of $\ket{\tilde{X}_0}_{(q,q)}$ due to the factor of $\frac{1}{\sqrt{2}}$ in Eq.~\eqref{eq:newdensitymatrix}.

\subsubsection{Conformal weight}\label{sec:ConformalWeight}

\begin{center}
	\begin{figure}[htb]
		\begin{center}
			\includegraphics[width=0.95\linewidth]{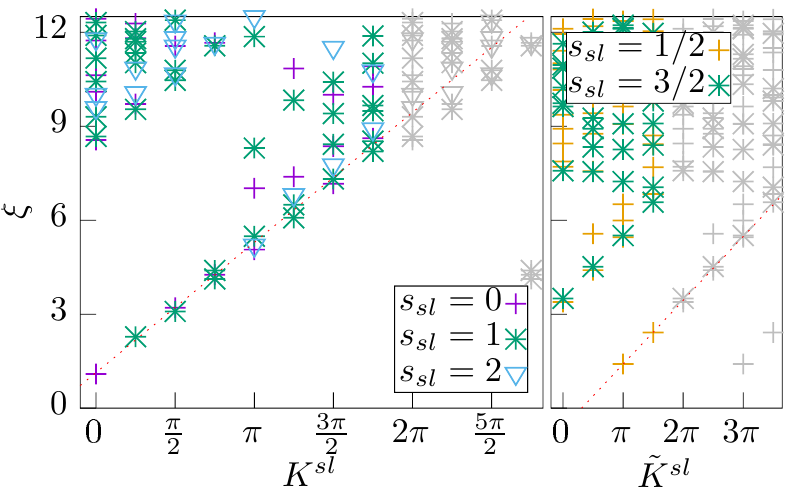}
		\end{center}
		\caption{Entanglement spectrum for the chiral PEPS with fixed virtual $\uone$ charge in (a) for $q=0$ corresponding to integer spin and in (b) for $q=1$ corresponding to half-integer spin. The dotted lines are linear fits obtained from the averaged multiplets in the lowest four and three levels for (a) and (b), respectively. We plotted parts of the second BZ in gray to show the chirality of the branches extending for more than a single BZ. The parameter values are $\lambda_1=0$, $\lambda_2 = \lambda_c =1$ and the system size is $N_v=8$.}
		\label{fig::ESUone} 
	\end{figure}
\end{center}

In Fig.~\ref{fig::ESUone} we display the ES derived from the virtual reduced density matrix Eq.~\eqref{SigmaLinSup} for the chiral PEPS on a cylinder of width $N_v = 8$ with parameters $\lambda_1=0$ and $\lambda_2 = \lambda_c = 1$ which is close to the point where the overlap between the states with charges $q=\pm 1$ is maximal. We used the maximal set of quantum numbers such that the ES is computed using the single-step (double-step) translation in the sector $q=0$ ($q=1$). The low-lying entanglement energies in either sector lie on a chiral branch where the dispersion velocity is nearly the same in both sectors and the counting of the $\sutwo$ multiplets is precisely that of the CFT $\mathfrak{su}(2)_1$ up to the first four levels~\cite{francesco2012conformal}. 

\begin{center}
 \begin{figure}[htb]
\begin{center}
\includegraphics[width=0.95\linewidth]{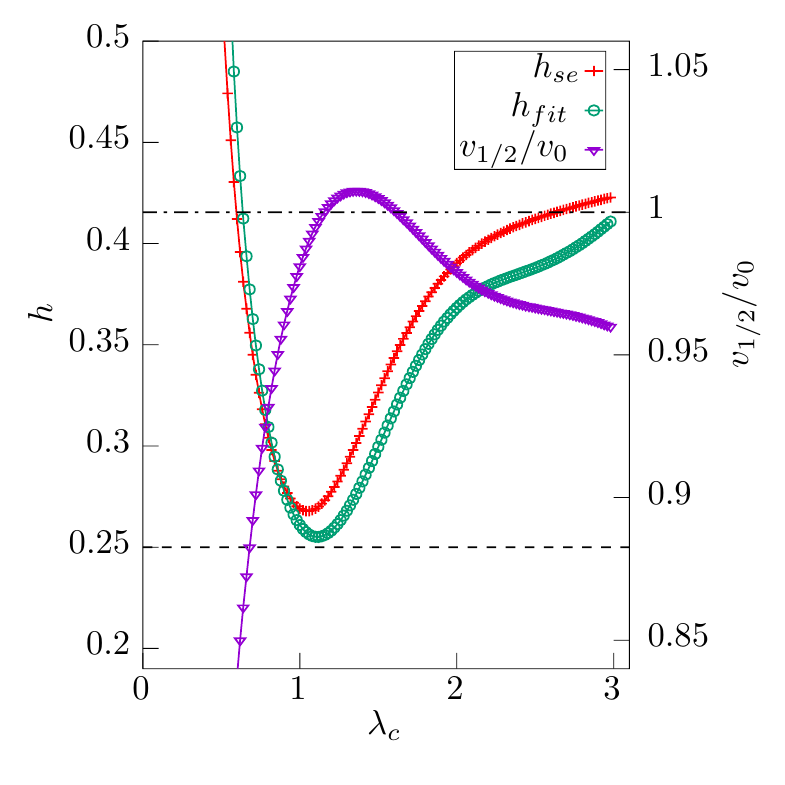}
\end{center}
\caption{Numerical estimates for the conformal weight of the spin-1/2 sector and the ratio of the half-integer and integer spin dispersion velocities as a function of $\lambda_c$. In the vicinity of $\lambda_c =1$, the conformal weight approaches the expected value of 1/4 and the dispersion velocities are approximately equal as required by conformal invariance. The expected values for these quantities are indicated by dashed lines. Here, the data was analyzed using a linear fit and the estimates for the conformal weight are obtained using Eqs.~\eqref{SimpleEstimate} and~\eqref{LinearFit}, respectively. The PEPS parameters are $\lambda_1 = 0$ and $\lambda_2 =1$ and the cylinder width is $N_v = 8$.}
\label{fig::ConformalWeight} 
\end{figure}
\end{center}

We studied different properties of the ES in order to quantify how closely it resembles the CFT $\mathfrak{su}(2)_1$ spectrum for different values of $\lambda_c$. Firstly, we computed the conformal weight of the half-integer spin sector using two different methods. On one hand, we used a simple estimate
\begin{equation}\label{SimpleEstimate}
h_{se}\approx\frac{\xi_0^{(1/2)} - \xi_0^{(0)}}{\xi_1^{(0)} - \xi_0^{(0)}}
\end{equation}  
where $\xi^{(s)}_i$ is the entanglement energy of the $i$\textsuperscript{th} level in the sector with spin $s=0$ or $s=1/2$. On the other hand, we approximated the average entanglement energy of the multiplets corresponding to the lowest four (three) levels of the chiral branch in the ES of the spin-0 (spin-$1/2$) sector using a linear fit with the offset $a_s$ and dispersion velocity $v_s$ as free parameters. The conformal weight is then given by
\begin{equation}\label{LinearFit}
h_{fit}\approx \frac{2}{v_0 + v_{1/2}}(a_{1/2}-a_0).
\end{equation}
Secondly, we compared the ratio $v_{1/2}/v_0$ of the dispersion velocities for the two sectors as obtained from the linear fits. Since a CFT has no mass scale, these velocities are equal for $\mathfrak{su}(2)_1$. A third measure for the chirality of the ES is given by the difference in entanglement energy between the second-lowest level at $K^{sl}= 0$ and the lowest state at momentum $K^{sl}= 7$. This difference should be positive if the chiral branch wraps more than once around the BZ. We observe that this criterion is fulfilled for $0.6\leq \lambda_c\leq 1.12$, showing that the correspondence to the CFT is closest inside this region. In Fig.~\ref{fig::ConformalWeight} we show the conformal weight and the ratio of the two dispersion velocities for $N_v=8$ as a function of $\lambda_c$. The conformal weight derived from either estimate has a clear minimum in the vicinity of $\lambda_c= 1$ with values of $h\approx 0.27$ for the simple estimate and $h\approx 0.25$ for the linear fit. This is in very good agreement with the value of $h=1/4$ expected for $\mathfrak{su}(2)_1$~\cite{francesco2012conformal}. For small and very large values of $\lambda_c$ the estimated conformal weight grows rapidly as the ES becomes gapped. This is expected as we showed in Sec.~\ref{sec:PEPSDef} that the PEPS is real both for $\lambda_c = 0$ approached in the limit $\lambda_c \ll \lambda_2$ and for $\lambda_2 = 0$ which up to an overall normalization is the state we obtain in the limit $\lambda_c \gg \lambda_2$. Moreover, the ratio of estimated dispersion velocities for the half-integer and integer spin sectors is very close to unity in the vicinity of $\lambda_c= 1$ whereas it decreases rapidly for $\lambda_c < 1$. All in all, for $0.8\leq \lambda_c \leq 1.12$ the estimated conformal weight is less than 0.3, the ratio of the two dispersion velocities deviates by less than $10\%$ from the expected value and the chiral branch in the integer spin sector wraps more than once around the BZ. We therefore conclude that the chiral PEPS in this parameter region possesses an ES whose low-lying levels correspond very closely to the spectrum of the chiral CFT $\mathfrak{su}(2)_1$ since it has the correct state counting, conformal weight and identical dispersion velocities in both sectors.

\section{Conclusion}

In this article we investigated the interplay of point group symmetry and translation symmetry for $\sutwo$ invariant PEPS. We showed that for half-integer physical spins it is not possible to simultaneously impose both translation invariance and point group symmetry at the level of the local tensors. Depending on which symmetry is imposed one obtains generically distinct PEPS that are related by the insertion of virtual $\mathbb{Z}_2$ flux strings. This understanding enabled us to explain the discrepancies between the spectrum of the chiral CFT $\mathfrak{su}(2)_1$ and the ES of the chiral spin liquid PEPS introduced in Ref.~\cite{Poilblanc2015PRB_chiralPEPS,Poilblanc2016PRB_chiralPEPS}. Moreover, we were able to identify a region of the parameter space where these discrepancies can be lifted due to a staggered virtual $\uone$ symmetry of the PEPS and presented numerical data establishing a correspondence between the ES and the CFT spectrum. Many questions remain open in the application of PEPS to chiral topologically ordered states. A crucial missing ingredient is an analytical understanding of the link between the symmetry structure of the local tensors and the CFT obtained in the entanglement spectrum. Such an understanding would permit the generalization of the PEPS we analyzed in this article to other edge CFTs. Another important direction would be the development of additional probes to detect the edge chirality and to identify the precise nature of the edge CFT. These would allow for more precise disambiguation of regimes with chiral and gapped entanglement spectra.

\begin{acknowledgments}

We are grateful to F.~Alet, J.~Dubail, D.~Poilblanc, G.~Sierra, H.-H.~Tu, F.~Verstraete and E.~Zohar for fruitful discussions. This work was supported by the European Union through the ERC Starting Grant WASCOSYS (No. 636201). AS acknowledges funding by the Alexander von Humboldt Foundation. 
\end{acknowledgments}

\appendix*
\section{\label{app:tensor}Tensor elements}

In this appendix we provide an explicit expression for the local projection map Eq.~\eqref{defA} of the chiral spin liquid PEPS. In contrast to Ref.~\cite{Poilblanc2015PRB_chiralPEPS,Poilblanc2016PRB_chiralPEPS} we use a representation where the local tensor transforms under the point group as $\mathbf{A}_1 +i \mathbf{A}_2$ rather than $\mathbf{B}_1 +i \mathbf{B}_2$. As explained in Sec.~\ref{sec:PEPSDef}, both representations for the local tensor define the same state. We denote by $\ket{0},\ket{1}$ the eigenstates of the spin-$\frac{1}{2}$ representation with $S^z$ eigenvalue $\pm \frac{1}{2}$ both on the physical and virtual legs whereas $\ket{2}$ corresponds to the virtual spin-0 state. In this basis, the non-vanishing tensor elements are given by
\begin{gather}
A^s_{s222} = A^s_{2s22} = A^s_{22s22} = A^s_{222s} = \lambda_1,\\
A^s_{s\sbar s2} = A^s_{\sbar s2s} = A^s_{s2s\sbar} = A^s_{2s\sbar s} = -2(-1)^s\lambda_2,\\
A^s_{ss\sbar 2} = A^s_{s\sbar 2s} = A^s_{\sbar 2ss} = A^s_{2s2\sbar} = (-1)^s[\lambda_2+i\lambda_c],\\
A^s_{\sbar ss2} = A^s_{ss2\sbar } = A^s_{s2\sbar s} = A^s_{2\sbar ss} = (-1)^s[\lambda_2-i\lambda_c],
\end{gather}
where $s=0,1$.

\end{document}